\documentclass[%
 reprint,
 amsmath,amssymb,
 aps,
]{revtex4-2}

\usepackage{graphicx}
\usepackage{dcolumn}
\usepackage{bm}

\usepackage{enumitem} 
\usepackage{subfigure}
\usepackage{float}
\usepackage{color}
\usepackage{ulem}
\usepackage{amsmath}

\newcommand\COMMENTED[1] {}

\begin{document}

\title{Correlation effects in magic-angle twisted bilayer graphene: \\
An auxiliary-field quantum Monte Carlo study}

\author{Zhi-Yu Xiao}
\affiliation{%
Department  of  Physics,  College  of  William  \&  Mary,  Williamsburg,  Virginia  23187,  USA
}%

\author{Shiwei Zhang}
\affiliation{%
Center  for  Computational  Quantum  Physics,  Flatiron  Institute,  New  York,  NY  10010,  USA
}%
\date{\today}

\begin{abstract}
Magic angle twisted bilayer graphene (MATBG) presents a fascinating platform for investigating the effects of electron interactions in topological flat bands. The Bistritzer-MacDonald (BM) model provides a simplified quantitative description of the flat bands. Introducing long-range Coulomb interactions leads to an interacting BM (IBM) Hamiltonian, a momentum-space continuum description which offers a very natural starting point for many-body studies of MATBG. Accurate and reliable many-body computations in the IBM model are 
challenging, however, and have been limited mostly to special fillings, or smaller lattice sizes. We employ state-of-the-art auxiliary-field quantum Monte Carlo (AFQMC) method to study the IBM model, which constrains the sign problem to enable accurate treatment of large system sizes. We determine ground-state properties and quantify errors compared to mean-field theory calculations.
Our calculations identify correlated metal states and their competition with the insulating Kramers inter-valley coherent state at both half-filling and charge neutrality. Additionally, we investigate one- and three-quarter fillings, and examine the effect of
many-body corrections beyond single Slater determinant solutions. We discuss the effect that details of the IBM 
Hamiltonian have on the results, including different forms of double-counting corrections, and the need to establish and precisely specify many-body Hamiltonians to allow more direct and quantitative comparisons with experiments in MATBG.
 
\end{abstract}

\pacs{Valid PACS appear here}

\maketitle

\section{\label{sec:level1}INTRODUCTION}
The interaction-driven insulating phase and superconducting phase discovered in magic-angle twisted bilayer graphene (MATBG) \cite{TBG_Cao2018_Ins, TBG_Cao2018_SC} are inspiring intense efforts, in both the theoretical and experimental domains, to understand the mechanism behind them \cite{BM_model, PhysRevLett.121.087001, PhysRevX_Wannier, TBG_CIS_Allen, TBG_Hubbard_Sun, PhysRevB.98.085435, PhysRevB.100.035448, TBG_Origin_of_Magic_Angles, PhysRevLett.122.246401, PhysRevLett.122.246402, Symmetry_of_TBG, skyrmions_topology_of_TBG, TBG_Bernevig_2021PRB_5, KSI_in_TBG, FlatBand_Utama2021, FlatBand_Lisi2021, TBG_Quantum_textures}. However, due to the interplay between
electron-electron interactions and flat bands with non-trivial topology, accurate, ab initio modeling of TABLG presents many challenges \cite{Symmetry_of_TBG, skyrmions_topology_of_TBG, TBG_Fragile_Topology, TBG_Fragile_Topology2, TBG_Bernevig_2021PRB_2}. The exploration of effective models for MATBG falls 
loosely 
into several 
categories: Hubbard-like models \cite{PhysRevX_Wannier, PhysRevX_Wannier2} based in real-space; 
the so-called interacting Bistritzer and MacDonald (BM) model
based in momentum space,
by including long-range electron-electron interactions
in the low-energy continuum BM model \cite{BM_model}; 
and more recently, a heavy fermion description \cite{Heavy_Fermion_Bernevig_2022_PRL, Heavy_Fermion_DMFT}.  

In this work, we take the second approach, and perform a systematic many-body study using the interacting BM (IBM) model.
Previous studies of the IBM model have sought to 
explain the interaction-driven insulating phase 
with perturbation theory \cite{Symmetry_of_TBG, TBG_Bernevig_2021PRB_5} and Hartree Fock (HF) approaches \cite{Symmetry_of_TBG, TBG_CIS_Allen, HF_PRB2020, HF_PRL2022, HF_PRR2021}. 
Many-body studies, including exact diagonalization (ED) \cite{TBG_Bernevig_2021PRB_5, ED_PRL_2021}, determinantal quantum Monte Carlo (DQMC) \cite{DQMC_2022_Johannes, DQMC_2021_CPL, DQMC_2021, DQMC_Huang_2024_TBG_Qfilling}, density matrix renormalization group (DMRG) \cite{DMRG_2020PRB, DMRG_2022PRB, DQMC_2021, DMRG_2021_Strain, DMRG_wang2022_Strain,DMRG_wang2022_Strain} and quantum chemistry methods \cite{QC_LinLin_faulstich2022interacting} have shown that the proper treatment of electron-electron interactions and their interplay with dispersion are indispensable for determining ground state properties. 
Much progress has been made in characterizing the 
properties of this model. 
Important obstacles remain, however, for 
establishing comparisons and
connections with experiments \cite{TBG_Cao2018_Ins, Polshyn2019, Lu2019, Saito2020, Stepanov2020, Saito2021, TBG_Quantum_textures}.

The obstacles are several fold. 
The relevant model parameters and electron fillings lie within the most difficult regime, and require very high resolution to capture accurately the interplay between band dispersion and electron-electron interactions. This is at the heart of strong correlation 
physics, and presents challenges of various forms to every computational method. Here the presence of 
topological bands and long-range interactions 
necessitates large system sizes in order to properly 
reach the thermodynamic limit. 
It has also become clear that 
the answers can be sensitive to 
model details 
including double counting corrections, the effect of remote bands, and accounting for strain, etc.  
The lack of accurate solutions to the model makes it more challenging to 
pinpoint and understand discrepancies with experimental observations. 

Here we tackle this problem 
by employing the phaseless auxiliary-field quantum Monte Carlo (AFQMC) method \cite{AFQMC_Zhang-Krakauer-2003-PRL} to investigate the ground state of the IBM model at integer fillings.
Our approach is based on the momentum space description and 
captures the non-trivial topology by retaining eight ``flat'' bands.
Long-range (screened) electron-electron interaction is treated explicitly and rigorously. 
The phaseless AFQMC method mitigates the sign/phase problem by a constrained path approximation
\cite{QMC_Zhang_Constrained_1997,lecturenotes-2019}; it enables simulations of large lattice sizes
(here calculations are performed up to $12\times 12$ cells), from which properties at the thermodynamic limit can be inferred. 
The accuracy of the contrained path 
approximation
(more precisely the phaseless approximation \cite{AFQMC_Zhang-Krakauer-2003-PRL} because of the long-range interaction) depends on the 
form of Hubbard-Stratonovich transformation.
The particular form to
decompose a long-range interaction written in momentum 
space has been applied in many 
AFQMC studies of solids \cite{Planewave-AFQMC-PhysRevB.75.245123}. A more general form to decompose 
a 4-index Coloumb interaction written in an arbitrary 
basis is also available 
\cite{Cholesky_dec,Hao_Some_recent_developments}, and has been 
applied for two decades in quantum chemistry \cite{Joonho_chemistry,transition_metals_Shee}.
The extensive benchmarks in these show that the phaseless AFQMC approach allows accurate treatment of 
correlated materials with long-range interactions.

Using this approach we perform a systematic and comprehensive exploration of the ground-state phase diagram of the IBM model. 
We focus on 
four integer fillings that are specified by the number of electrons per ${\mathbf k}$-point $n_f$ (charge neutrality $n_f=4$, three quarter $n_f=3$, half-filling $n_f=2$, and quarter-filling $n_f=1$). Our study scans a wide range of tunneling ratio $\kappa$ that captures the lattice relaxation between the intra-sublattice 
(AA)
and inter-sublattice (AB$\&$BA) regions. 

At charge neutrality and half-filling, a transition from correlated insulating states (i.e., Kramers inter-valley coherent, K-IVC) to 
metallic states is observed as the 
tunneling ratio $\kappa$ is increased. 
This transition is characterized by the correlation energy, charge gap,
and related order parameters. Our results indicate a loss of K-IVC long-range order which contributes to the insulator-metal transition.  At quarter-filling, a transition from the correlated insulating state to a semi-metallic state is observed
as a function of 
tunneling ratio $\kappa$. 
This transition is characterized by charge gap, and C$_{2}$T symmetry. (Here, C$_{2}$T indicates 180-degree in-plane rotation symmetry C$_{2}$ with time reversal T.)
The AFQMC results confirm that the ground states in both phases can be well described by a single determinant (i.e., solution of HF), 
in agreement with previous works using quantum chemistry
methods 
(treating no valley degree of freedom) \cite{QC_LinLin_faulstich2022interacting}. 
We also find 
a degenerate 
state with correlations between two valleys in the semi-metallic regime.

These results, in comparison with experimental observations, will also shed light on the modeling of MATBG. Although much needs to be done in this model, including 
fractional fillings, the systematic and accurate results 
at integer filling provide an important step in determining 
what additional ingredients, if any, will need to be considered to quantitatively describe MATBG. 

The rest of the paper is organized as follows. In Sec.~\ref{sec:IBM_model}, we briefly introduce the IBM model. In Sec.~\ref{sec:AFQMC}, we introduce our AFQMC 
algorithm for studying Coulomb interaction with the topological band and its typical application in the IBM model. In Sec.~\ref{sec:results}, we describe the zero-temperature phase diagram of the IBM model 
at integer fillings. Finally, in Sec.~\ref{sec:summary}, we 
discuss connections with experiments 
comment on possible future work. 

\section{\label{sec:IBM_model}IBM model}
Our model consists of the continuum BM model $\hat{H}_{BM}$, 
which describes the moire bands 
at charge neutrality, and long-range screened Coulomb interactions $\hat{V}$. We briefly introduce IBM model here while elaborating more details in App.~\ref{sec:APPENDIX_IBM}. 

The continuum BM model $\hat{H}_{BM}$ is a tight-binding model that describes electrons hopping between $2p_z$ orbitals of carbon atoms in twisted bilayer graphene. By taking Bloch vector $\boldsymbol{p}$, sub-lattice $\sigma$, and layer $\mu$, spin $s$ to identify the Bloch state (see Eq.~\ref{eq:Bloch_State}), the continuum BM model can be written as:
\begin{equation}
\begin{aligned}
\hat{H}_{BM}&=\sum_{\boldsymbol{p}}T^{\textup{intra}}_{\boldsymbol{p},\boldsymbol{p}}c^\dagger_{\boldsymbol{p}}c_{\boldsymbol{p}} + T^{\textup{inter}}_{\boldsymbol{p}',\boldsymbol{p}}c^\dagger_{\boldsymbol{p}'}c_{\boldsymbol{p}}\\
T^{\textup{intra}}_{\boldsymbol{p},\boldsymbol{p}}&=\hbar\upsilon _F(p_x\sigma_x\tau_z + p_y\sigma_y)\\
T^{\textup{inter}}_{\boldsymbol{p}',\boldsymbol{p}}&=\frac{1}{2}\sum_{l=0,1,2}\delta_{\boldsymbol{p}',\boldsymbol{p}-\tau_z \boldsymbol{q}_l}(\mu_x-i\mu_y)T_l + \textup{h.c.}\\
T_l&=\omega_0e^{-i\theta \sigma_z \mu_z \tau_z/2} + \omega_1e^{2\pi i l \sigma_z \tau_z/3} \sigma_x e^{-2\pi i l \sigma_z \tau_z/3}
\label{Equ.BM}
\end{aligned}
\end{equation}
where $\theta = 1.05^{\circ}$ is the twist angle,  $\nu_F=2700*\frac{3a_0}{2\hbar}$ meV is the Dirac velocity, and $\omega_0$, $\omega_1$
describes lattice relaxation between A-A and A-B stacks, respectively. Transition momentum $\boldsymbol{q}_l=R(2\pi l/3)\boldsymbol{q}_0$, where $R(\psi)=e^{i\psi \sigma_y}$ is a rotation matrix. $\boldsymbol{q}_0=\frac{8\pi \textup{sin}(\theta/2)}{3\sqrt{3}a_0}(0,-1)^T$ and $a_0=1.42 \AA$ is the intra-layer distance between carbon atoms. 

As two graphene layers stack with a twisted angle $\theta$, the Brillouin zones (BZ) of mono-layer graphene rotate with angle $\theta$ related to the other layer, and this rotation leads to periodic moire pattern and folds BZ to mini-BZ (MBZ) \cite{BM_model}. That is we use $\varepsilon _{\boldsymbol{k},i}$ and $|\varphi_{\boldsymbol{k},i} \rangle$ to specify each band $n=0,...,N_G-1$ in MBZ:
\begin{equation}
\hat{H}_{BM}|\varphi_{\boldsymbol{k},i} \rangle=\varepsilon _{\boldsymbol{k},i}|\varphi_{\boldsymbol{k},i} \rangle
\end{equation}
with $i=(n,\tau,s)$ to label bands $n$, valley $\tau$, and spin $s$. $N_G$ is truncated for numerical implementation.

\begin{figure}[htbp]
\includegraphics[scale = 0.2]{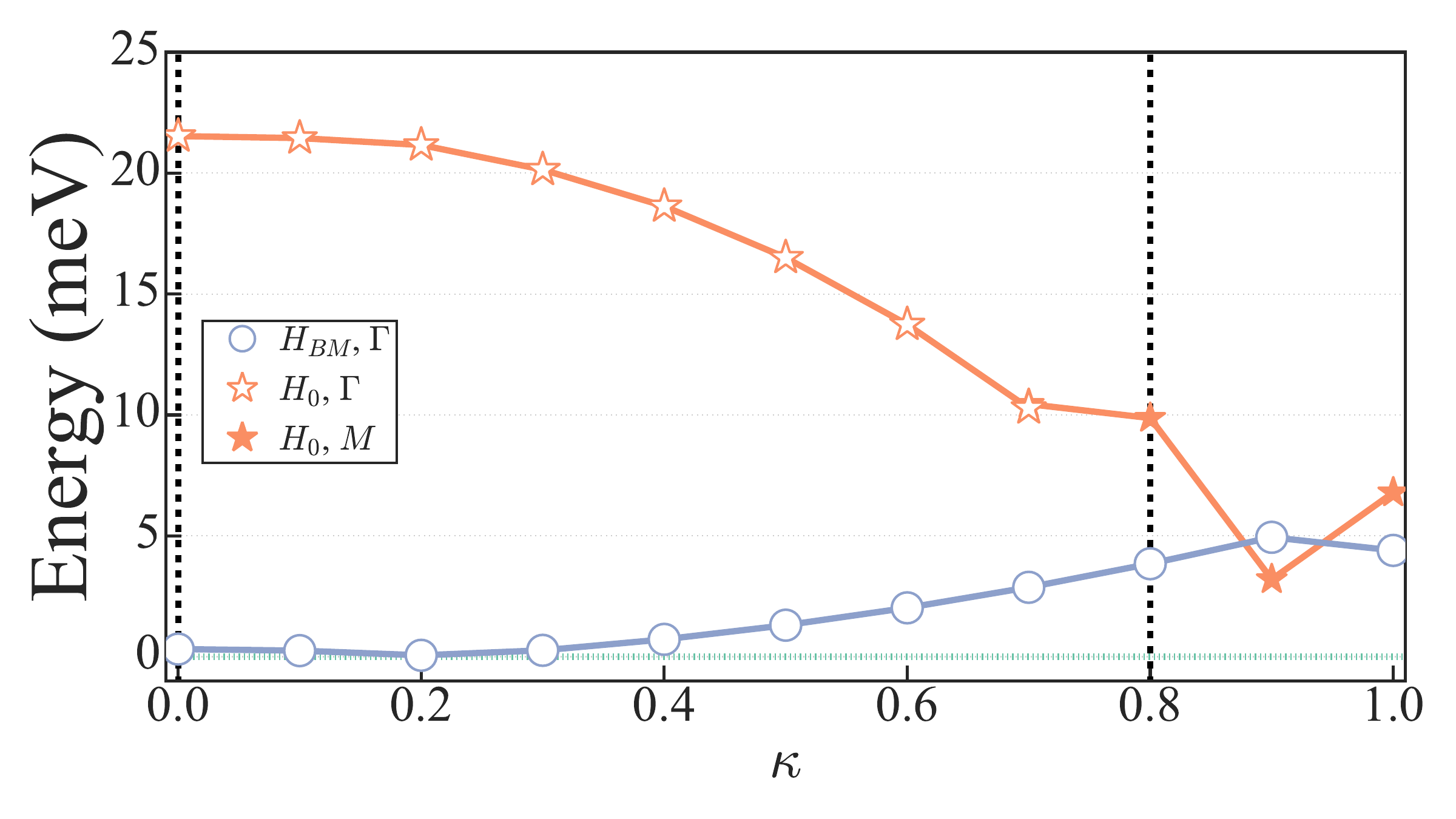} 
\includegraphics[scale = 0.175]{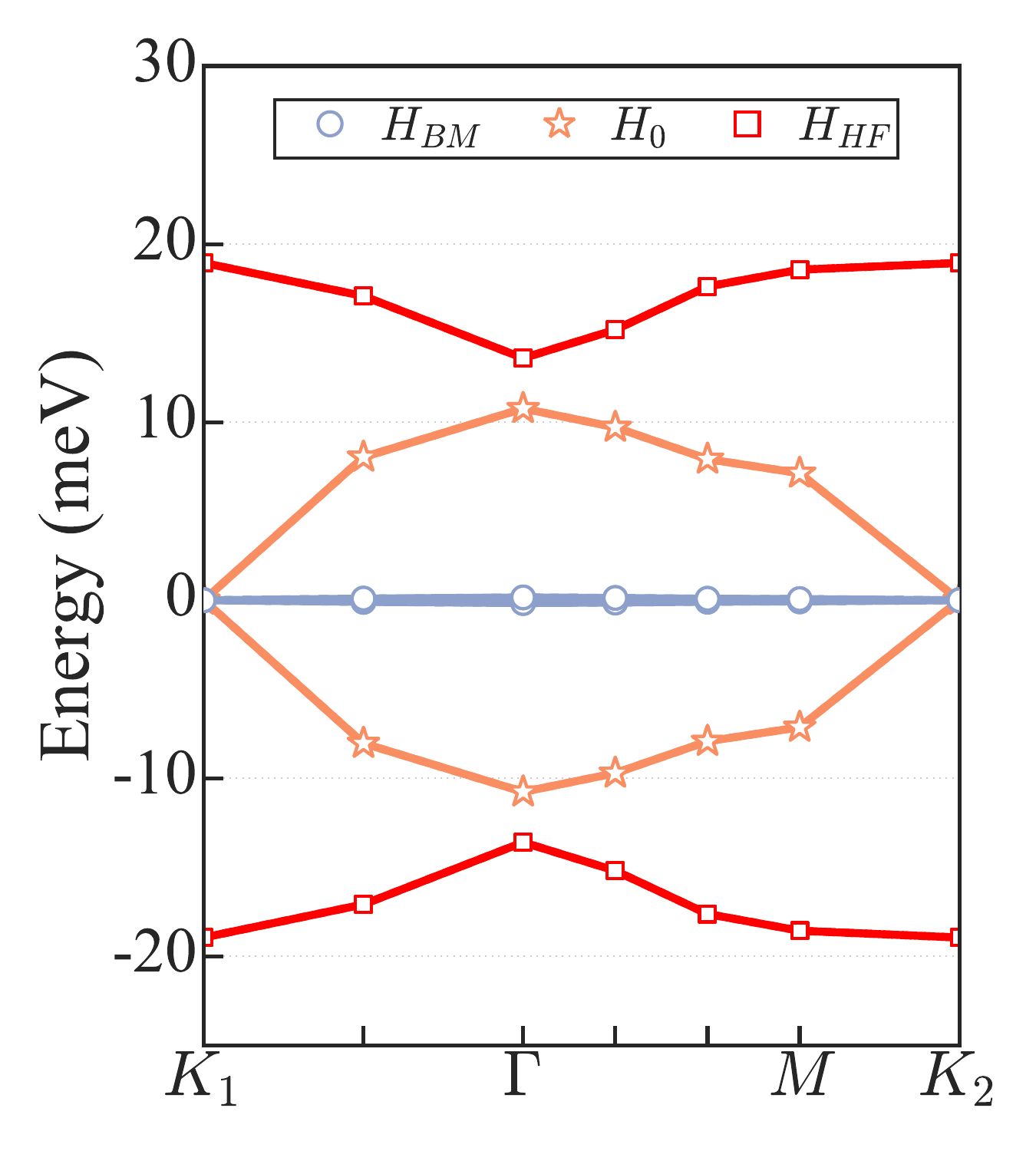}  
\includegraphics[scale = 0.175]{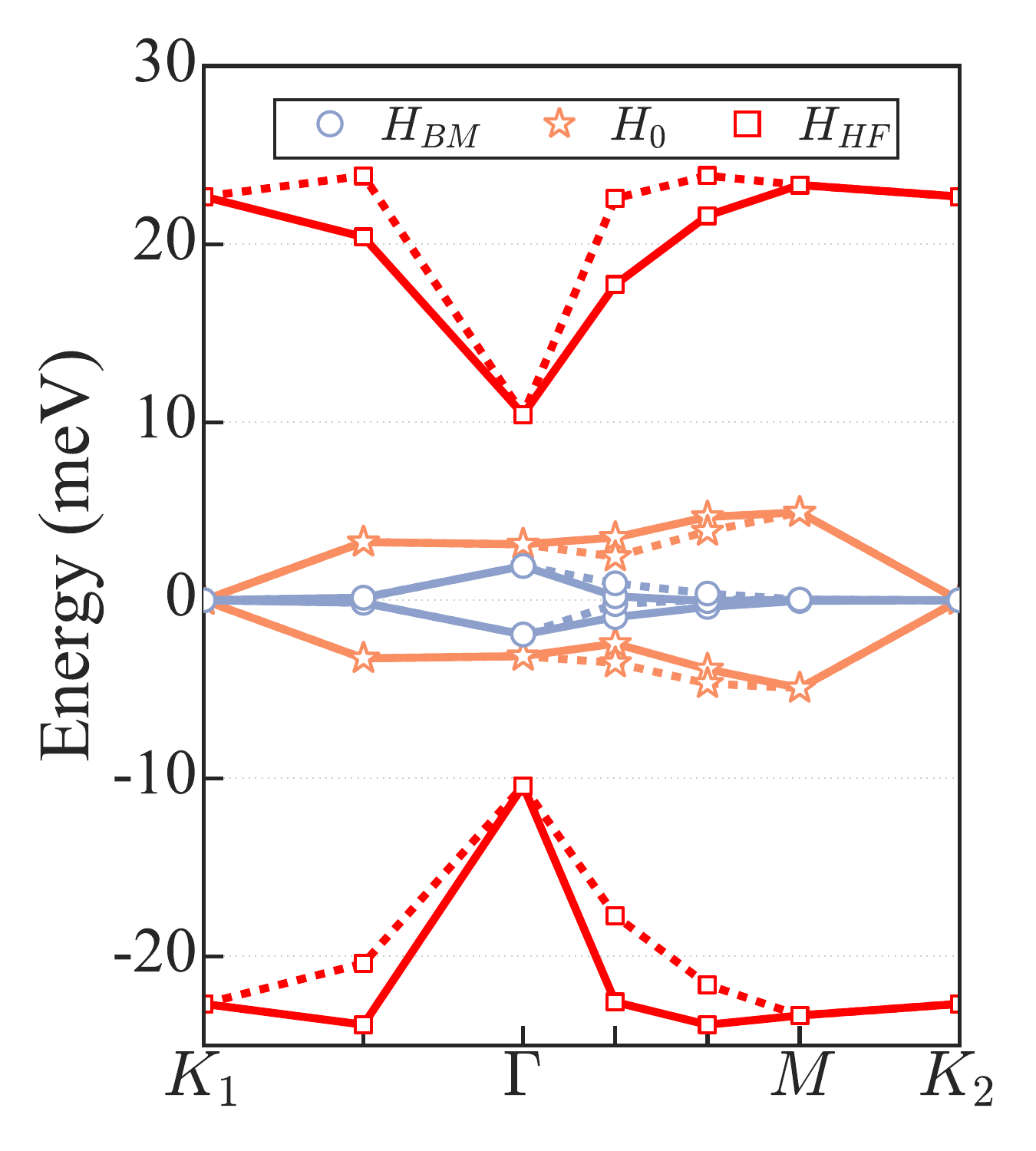} 
\caption{(Top) 
Comparison of bandwidths between $\hat H_{BM}$ and $\hat H_{0}$ versus 
$\kappa$ in a $6 \times 6$ system. $\Gamma$ and $M$  label the ${\mathbf k}$-point where the largest gap is observed. (Bottom) Band structures of $\hat H_0$, $\hat H_{BM}$ and $\hat H_{HF}$ for $\kappa=0.0$ and $\kappa=0.8$. $\hat H_{HF}$ indicates the IBM model band structure obtained from Hartree-Fock (HF). Solid and dashed lines specify the two valleys respectively (degenerate at $\kappa=0$). 
} 
\label{Fig.H_bandstructure} 
\end{figure}

We mention here that a correction
term is included
in addition to 
the BM model Hamiltonian $\hat{H}_{BM}$.
The additional correction term, $[\hat{V}]_{\phi_{BM}}$, 
is a mean-field approximation of the Coulomb interactions $\hat{V}$ with respect to the ground state $\phi_{BM}$ of the BM model at charge neutrality. This correction is adopted to remove or counter 
the double counting of the interaction effect \cite{Symmetry_of_TBG, HF_PRR2021, DQMC_2022_Johannes}, as parameters used in the BM model are 
based on first-principle calculations \cite{BM_model} which already account for 
the effect of Coulomb interactions at the density-functional theory level. 
There are somewhat different ways to 
treat the correction 
\cite{QC_LinLin_faulstich2022interacting, DQMC_2021_CPL, DQMC_2021, DQMC_Huang_2024_TBG_Qfilling}, and the results turn out to be 
quite sensitive to the details of the correction.
As we illustrate in Fig.~\ref{Fig.H_bandstructure},
the band-structure of the dispersion part of the Hamiltonian, $\hat H_0\equiv\hat{H}_{BM} - [\hat{V}]_{\phi_{BM}}$,
can be significant different from that of $\hat{H}_{BM}$.

It is also important to mention that the model is parameterized by the ratio $\kappa=\omega_0/\omega_1$, which describes the lattice relaxation between AA regions (characterized by $\omega_0$) and AB$\&$BA regions (characterized by $\omega_1$). This ratio $\kappa$ is responsible for the emergence of flat bands \cite{TBG_Origin_of_Magic_Angles} and the ground state symmetry \cite{skyrmions_topology_of_TBG, Symmetry_of_TBG}. To explore the effects of lattice relaxation in TBG, 
we fix $\omega_1 = 105$meV and study  $\kappa \in (0, 1)$, which covers 
the entire range from 
the chiral limit ($\kappa=0$) through 
the more realistic case ($\kappa \sim 0.8$).

The IBM model is built based on eight ``flat" bands (i.e., two bands with two valleys and two spins that are closest to the Fermi surface ) from 
the BM model. 
Effect from the remote bands are neglected. 
The Hamiltonian of our IBM model is
\begin{equation}
\begin{aligned}
&\hat{H}_{IBM}=\hat{H}_0(\kappa) + \hat{V}\\
&=\sum_{\boldsymbol{k},i}\varepsilon _{\boldsymbol{k},i}c^\dagger_{\boldsymbol{k},i}c_{\boldsymbol{k},i} - \sum_{\boldsymbol{k},i,j}[V]_{\boldsymbol{k},i;\boldsymbol{k},j}c^\dagger_{\boldsymbol{k},i}c_{\boldsymbol{k},j} + \frac{1}{2}\sum_{\boldsymbol{q}} \bar{\rho}_{\boldsymbol{q}} V_{\boldsymbol{q}} \bar{\rho}_{-\boldsymbol{q}}.
\end{aligned}
\label{Equ.IBM}
\end{equation}
where $\boldsymbol{k}$ is on a hexagonal lattice 
with size $L$ to discretize the first 
MBZ (see Fig.~\ref{Fig.lattice}), $\bar{\rho}_{\boldsymbol{q}}$ is the density operator at momentum $\boldsymbol{q}=\boldsymbol{k}-\boldsymbol{k}'$ with a background shift \cite{Hao_Some_recent_developments}, and $V_{\boldsymbol{q}}$ is the Fourier transformed Coulomb potential. The transfer momentum  $\boldsymbol{q}$ remains unrestricted, but is truncated  for numerical implementation. The Coulomb potential is two-gate screened:
\begin{equation}
V_{\boldsymbol{q}}=\frac{e^2}{2\epsilon \epsilon _0}\frac{1}{A}\frac{\textup{tanh}(|\boldsymbol{q}|d)}{|\boldsymbol{q}|}.
\end{equation}
which is the Fourier-transformed Coulomb potential with the distance $d=20$ nm between two layers and relative permittivity $\epsilon=10$ for vacuum permittivity $\epsilon_0$. We comment on the choice of relative permittivity in App.~\ref{sec:APPENDIX_parameter}. $A = N_M A_0$ is the total area of the system with $A_0=\sqrt{2}a^2_M/2$, $a_M=\sqrt{3}a_0/2 \textup{sin}\frac{\theta}{2}=13.4$ nm and $N_M$ is the number of moire unite cells of this system.

\section{\label{sec:AFQMC}AFQMC with phaseless constraint}
The IBM model is a much simplified model, 
which 
still contains both a non-trivial band-structure, 
including corrections, and long-range Coulomb interaction. 
The phaseless AFQMC \cite{AFQMC_Zhang-Krakauer-2003-PRL}
has the capacity to study such a general Hamiltonian while 
controlling the sign problem to restore
low-polynomial computational scaling 
with lattice sizes. To give 
a brief self-contained description, we adopt the notation
in Ref.~\cite{Hao_Some_recent_developments} and start from the Hamiltonian we treat, written 
in ``Monte Carlo form":
\begin{equation}
\hat{H}= \hat{T}+\frac{1}{2}\sum^\Gamma_\gamma \hat{L}_\gamma ^2 
\label{eq:H_MC}
\end{equation}
where $\hat{T}$ and $\hat{L}_\gamma$ denote sets of one-body operators whose matrix elements are explicitly specified in the chosen basis. To study the IBM model, we take $\hat{L}_{\boldsymbol{q}} = \sqrt{\frac{V_{\boldsymbol{q}}}{2}}(\bar{\rho}_{\boldsymbol{q}} + \bar{\rho}_{\boldsymbol{-q}})$ and $\hat{L}'_{\boldsymbol{q}} = i\sqrt{\frac{V_{\boldsymbol{q}}}{2}}(\bar{\rho}_{\boldsymbol{q}} - \bar{\rho}_{\boldsymbol{-q}})$ \cite{Planewave-AFQMC-PhysRevB.75.245123}.

AFQMC performs 
ground-state calculations through the imaginary-time projection:
\begin{equation}
\begin{aligned}
|\Psi_{0}\rangle &\propto \underset{\beta \to \infty}{\lim} e^{-\beta\hat{H}}|\phi_I\rangle \\
&\approx  (e^{-\tau\hat{H}})^n|\phi_I\rangle\\
&\approx \int d\{\textbf{x}\} \prod^n_{i=1}
p(\textbf{x}_i)
\hat{B}(\textbf{x}_i) |\phi_I\rangle\,,
\end{aligned}
\label{eq:projection}
\end{equation}
which propagates the given initial state $|\phi_I\rangle$ to the ground state $|\Psi_{0}\rangle$ of the Hamiltonian $\hat{H}$,
if the overlap $\langle\Psi_{0}|\phi_I\rangle$ is non-zero. The second line of Eq.~(\ref{eq:projection}) is obtained by using the Suzuki-Trotter decomposition \cite{Suzuki,Trotter} to break the imaginary-time evolution operator into time slices with $n= \beta/\tau $. The third line 
is then obtained by applying the Hubbard-Stratonovich transformation \cite{HS_transformation_H, HS_transformation_S} to decouple the short-time propagator in each time slice with probability
\begin{equation}
p(\textbf{x})= \prod_\gamma \frac{1}{\sqrt{2\pi}}e^{-x_\gamma^2/2},
\end{equation}
and the one-body propagator 
\begin{equation}
\hat{B}(\textbf{x})=e^{-\tau\hat{T}/2}e^{\sum_\gamma x_\gamma\sqrt{-\tau}\hat{L}_\gamma }e^{-\tau\hat{T}/2}\,,
\label{eq:HS-B}
\end{equation}
where $\textbf{x}$ indicates a series of auxiliary-fields (AFs):  $\textbf{x}=\{x_\gamma\}$ 
($x_\gamma \in \mathbb{R}$). Without loss of generality, we can take $|\phi_I\rangle$ as a single determinant. Since a one-body propagator propagates a single determinant to another single determinant: $|\phi'\rangle = \hat{B}(\textbf{x})|\phi\rangle$, 
the integration in Eq.~(\ref{eq:projection}) can then 
be performed by Monte Carlo, mapping the sampling of 
$\{x_\gamma\}$ into random walks in the manifold of Slater determinants along the direction of imaginary time evolution \cite{lecturenotes-2019}.

Because of fermion antisymmetry, the Monte Carlo sampling of Eq.~(\ref{eq:projection}) in general suffers from 
a phase problem \cite{AFQMC_Zhang-Krakauer-2003-PRL}. The phase problem in AFQMC can be understood by considering \cite{lecturenotes-2019} a single determinant, through random chance in the time evolution, 
becoming perpendicular to the ground state, 
after which propagation of this determinant effectively contributes nothing but statistical fluctuations to the ground state estimation. Since the number of such determinants in general grows exponentially with projection time, the signal of Monte Carlo sampling will eventually be overwhelmed 
by statistic noise.  Eliminating these determinants that are perpendicular to the ground state is an exact condition which removes the sign or phase problem \cite{QMC_Zhang_Constrained_1997, AFQMC_Zhang-Krakauer-2003-PRL}. To implement this condition, AFQMC uses a trial wave function $|\Psi_T\rangle$ to constrain the random walk paths. For the general form of interaction, 
the orbitals in the Slater determinant become complex,
effectively introducing an overall gauge degree of 
freedom that is redundant in the projection.
The constraint in this case is a phaseless approximation that projects the weight of each Slater determinant sample 
to the real axis 
in the complex plane \cite{Hao_Some_recent_developments}. If the trial wave function is the exact ground state, then this constraint is unbiased. Many community benchmark studies have shown that AFQMC with constraint is highly accurate, both in model systems and in real materials \cite{simons_material_2020, Mario_hydrogen_chain, Hubbard_benchmark_2015}.

To further validate 
the accuracy and robustness of the phaseless AFQMC approach in this study,
we have taken several additional measures. 
In the IBM model different
low-lying states of different symmetry characters can be (nearly) degenerate. 
 The AFQMC method we use is in general much less 
 prone to ergodicity problems in the sampling, because it
 takes the form of many (almost independent) streams of 
 branching random walks. Still, we perform multiple 
AFQMC calculations using different low-lying HF states of different symmetry as trial/initial wave function.
The final energy is compared to gauge the preferred symmetry of the many-body ground state.
(Although the ground-state energy computed by AFQMC 
is not strictly variational \cite{issue-CPMC-1999},
the energy is very accurate, and we only draw conclusions
about the ordering of the states when the energy separation is much more significant than the expected resolution 
of the phaseless approximation.)
For the study of $n_f=2,4$,  C$_2$T-symmetry-breaking states and K-IVC states are barely distinguishable in energy around the chiral limit.
In these cases, we apply yet another step, using a constraint release method, AFQMC with Metropolis release constraint (AFQMC-MRC) \cite{zxiao_MRC_2023},
to minimize the constraint bias and identify the actual symmetry of the ground state. Further details are given in the App.~\ref{sec:APPENDIX_Low_lying_states}.

To summarize, in this work, we employ AFQMC to systematically investigate the IBM model. 
Compared with previous quantum Monte Carlo work \cite{DQMC_2022_Johannes, DQMC_2021_CPL, DQMC_Huang_2024_TBG_Qfilling}, 
this method is not limited to special fillings, high temperatures, or small system sizes. 
It  
allows us to access 
the whole range of $\kappa$ to determine the complete phase diagram (with the largest lattice up to $12\times 12$), 
while treating both valley and spin degrees of freedom. 

\section{\label{sec:results}Results}

In this section, we describe our numerical results 
for the ground state of the IBM model from
AFQMC, 
divided into several subsections 
by filling fractions: 
charge neutrality ($n_f=4$) and 
half-filling ($n_f=2$), quarter-filling ($n_f=1$), and three-quarter-filling ($n_f=3$).
The 
tunneling ratio $\kappa$ for realistic cases varies sample by sample and can 
affect the ground state behaviors. To present a systematic study 
and facilitate 
comparison with experiments, this work scans 
$\kappa$ from $0$ to $1$.
At the 
flat band limit, analytical studies \cite{skyrmions_topology_of_TBG, Symmetry_of_TBG} indicate that 
the candidate 
ground states are correlated insulating states: quantum anomalous Hall (QAH)
state, valley Hall (VH) state and 
K-IVC state for fillings $n_f=4$ and $2$, 
and QAH 
for filling $n_f=1$.
HF results 
(on the IBM model but with no double counting correction)
\cite{HF_PRB2020} 
support the above analysis and suggest the ground state at $n_f=3$ filling is a mixture of the states 
at 
$n_f=4,2$ and $n_f=1$ fillings. 
As discussed in the last paragraph of Sec.~III, we can use each of the candidate symmetry states above as the trial wave function in AFQMC. 
In this sense, our calculations can be viewed as using an unbiased (or at least less biased than other possibilities)
method, namely AFQMC, to judge the validity of the different 
candidates as the many-body ground state. 

Our results show that the HF solution is
in some cases quite accurate, while in other situations qualitatively incorrect. 
It is worth emphasizing that, while 
a weakly interacting system is often well described
by a single determinant solution from a mean-field 
calculation, the reverse is not 
always true. 
That is, a system which can be described well by a single Slater determinant 
is not 
necessarily weakly interacting. 
An example of this is the Wigner
crystal state in the electron gas at the low-density limit, an essentially classical state whose energy can be 
well approximated by that of a single-determinant 
solution in the spirit of unrestricted HF, but which 
is a strongly interacting state, 
with the interaction 
energy dominating 
the kinetic (zero-point) energy. 
In the IBM, the system is often more analogous to 
this situation when the HF solution is accurate.
We quantify the errors from mean-field theory calculations throughout the parameter regimes.

\subsection{\label{sec:CN_HF}Breaking of K-IVC insulating state at $n_f=4$ and $n_f=2$}

At $n_f=4$ and $n_f=2$ fillings, we observe a transition from a metallic state to a K-IVC insulating state.
At large $\kappa$ ($\kappa > \sim 0.8$), we find 
the ground states are K-IVC states characterized by a spontaneous hybridization between the two valleys that preserves an effective time-reversal symmetry.
The K-IVC state occupies both spins at charge neutrality and is polarized at one spin at half-filling. 
This is also the picture from the HF solution, but quantitative discrepancies
are seen between HF and AFQMC.
In contrast, at small $\kappa$ (i.e., $\kappa < \sim 0.2$)
we observe qualitative differences from the HF solution. 
The system is characterized by much larger correlation energy, a loss of long range K-IVC order, and vanishing charge gap. Below we first show our observation of this insulator-metal transition through the ground-state energy, including correlation energy, followed by
results on the charge gap. Then we study the K-IVC order parameter and band 
occupancy to complete this discussion.

\begin{figure}[htbp]
\includegraphics[scale = 0.2]{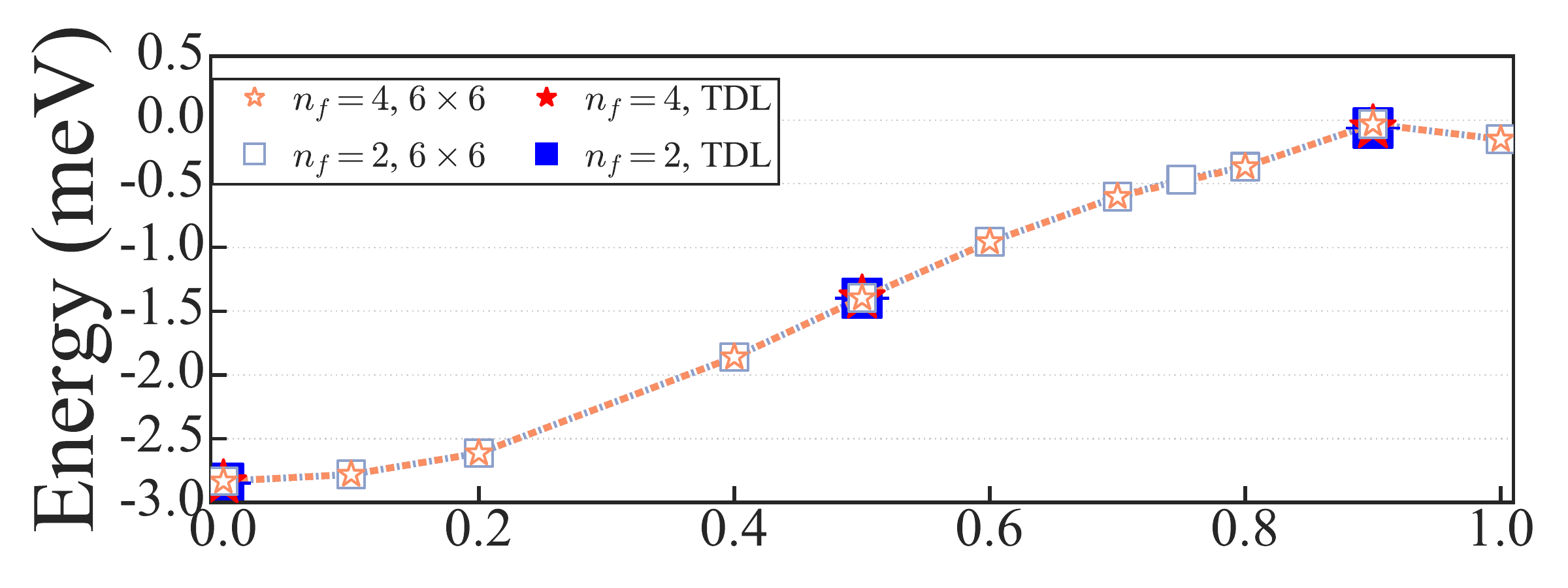}
\includegraphics[scale = 0.2]{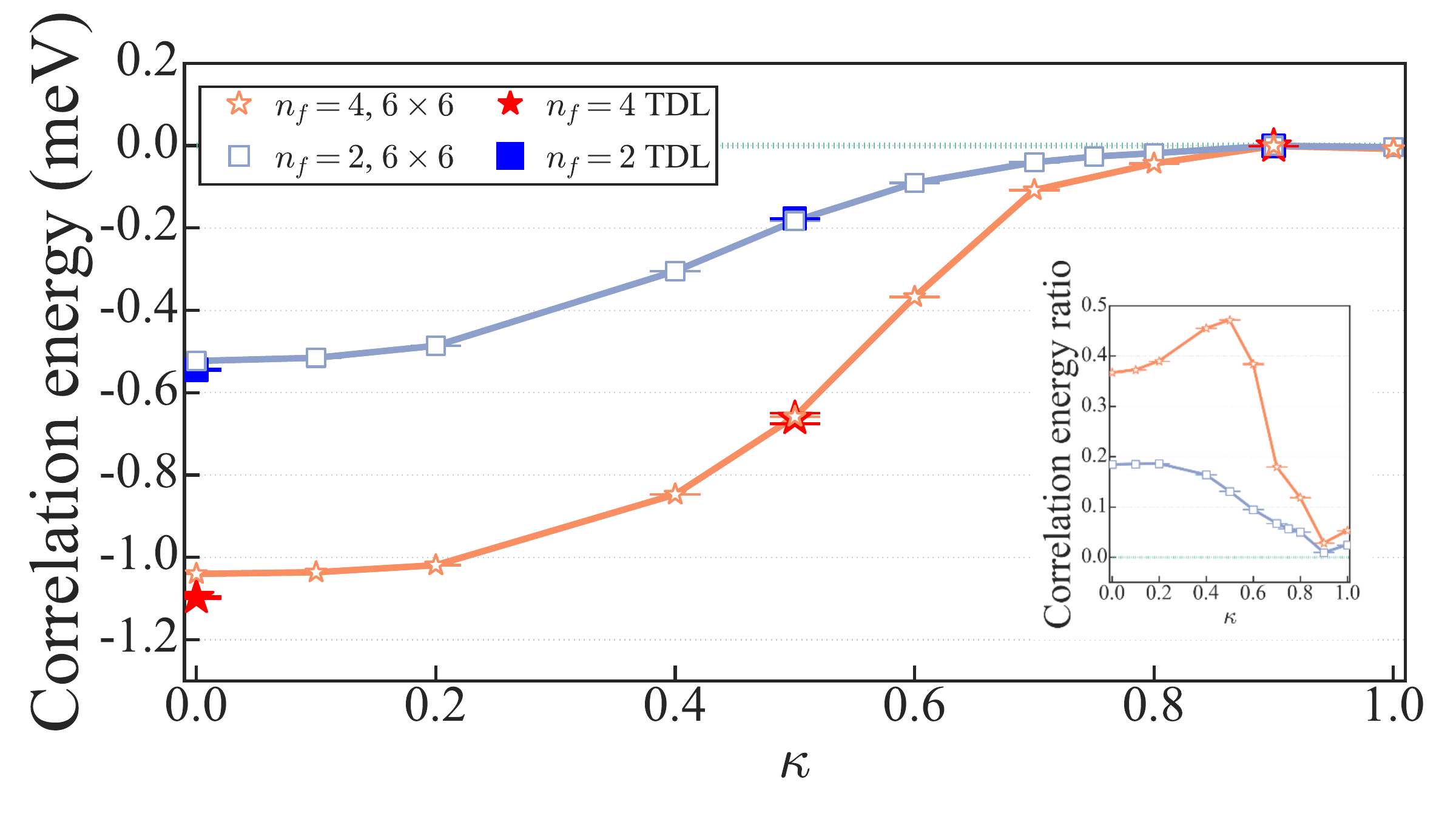}  
\caption{
Ground-state energies and correlation energies.
The top panel shows the HF energy per electron, 
$E_{\rm HF}$, as a function of $\kappa$,
for both $n_f=4$ and $n_f=2$. 
The bottom panel shows the corresponding
correlation energies, $E_c$. Most results are obtained from a $6\times 6$ 
lattice in momentum space, but at three values of $\kappa$ results are also shown 
for the thermodynamic limit (TDL),
which is obtained via extrapolation 
with $6\times 6$, $9\times 9$, and $12\times 12$ lattices.
The inset in the bottom panel shows the correlation energy relative to the total, $|E_c|/E_{\rm HF}$.
} 
\label{Fig.CorrelationEnergy_CN+HF} 
\end{figure}

In Fig.~\ref{Fig.CorrelationEnergy_CN+HF}, we show the HF energy per electron, $E_{\rm HF}$, 
and the  
AFQMC correction over HF  (i.e., the correlation energy) 
versus  $\kappa$.
The correlation energy is an important 
metric widely adopted in electronic structure 
calculations and quantum chemistry, 
defined as
the  difference between the true many-body 
ground-state energy and that from 
HF, which in this case is $E_c\equiv E_{\rm AFQMC}-E_{\rm HF}$, where
$E_{\rm AFQMC}$ is the ground-state energy per electron from AFQMC (calculated  
by the mixed estimator \cite{AFQMC_Zhang-Krakauer-2003-PRL,Hao_Some_recent_developments},
with the HF state as the trial wave function). For $n_f=2$, HF yields
a spin-polarized K-IVC state; for $n_f=4$, it finds 
a superposition 
of two spin-polarized K-IVC states with opposite spins and 
no correlation between them. 
The K-IVC states at the two fillings have exactly the same HF energy per electrons,
since the mean-field background of Coulomb interaction is removed 
in Eq.~(\ref{Equ.IBM}) by shifting the charge-density operators according to the chemical potential at
the corresponding 
filling $n_f$.
In the chiral limit ($\kappa=0$), HF 
yields an emergent C$_{2}$T-symmetry-breaking (QAH, VH) state 
that preserves $U_V(1)$ symmetry at both fillings and is degenerate with the K-IVC state. 
Both the K-IVC and the C$_{2}$T-symmetry-breaking states are correlated insulating states. Details of the above analysis are elaborated in App.~\ref{sec:APPENDIX_IBM_symmetry}. The above observation agrees with previous analytical studies \cite{skyrmions_topology_of_TBG, Symmetry_of_TBG} and the numerical HF  
results \cite{HF_PRB2020}. To simplify our studies, we focus on the K-IVC state here, 
but briefly discuss 
the close competition 
between it and the C$_{2}$T-symmetry-breaking state 
in App.~\ref{sec:APPENDIX_Low_lying_states}.

The amount of correlation energy $|E_c|$ grows 
as $\kappa$ decreases, as seen 
in the bottom panel of Fig.~\ref{Fig.CorrelationEnergy_CN+HF}.
Beyond $\kappa=0.9$ virtually no correlation energy is present, 
making the HF description essentially exact. 
At lower $\kappa$ many-body correlations make significant 
corrections to the HF results, which 
could render perturbative analysis 
more problematic. 
Interestingly, although the correlation energy magnitude
decreases monotonically with $\kappa$, 
the percentage of the correlation energy relative to the total 
ground-state energy (here measured with
respect to the HF energy) is not monotonic, reaching a maximum at intermediate $\kappa$ for $n_f=4$, 
as seen in the inset of the bottom panel.
We also observe a strong manifestation of 
correlation effects in the separation of the $n_f=4$ and  $n_f=2$ curves,  in contrast with 
previous analytical predictions \cite{skyrmions_topology_of_TBG, Symmetry_of_TBG} around the ``flat" band limit. Correlation between different
spins leads to a dramatic lowering of 
the $n_f=4$ correlation energy 
from the spin-polarized state. 

\begin{figure}[htbp]
\includegraphics[scale = 0.2]{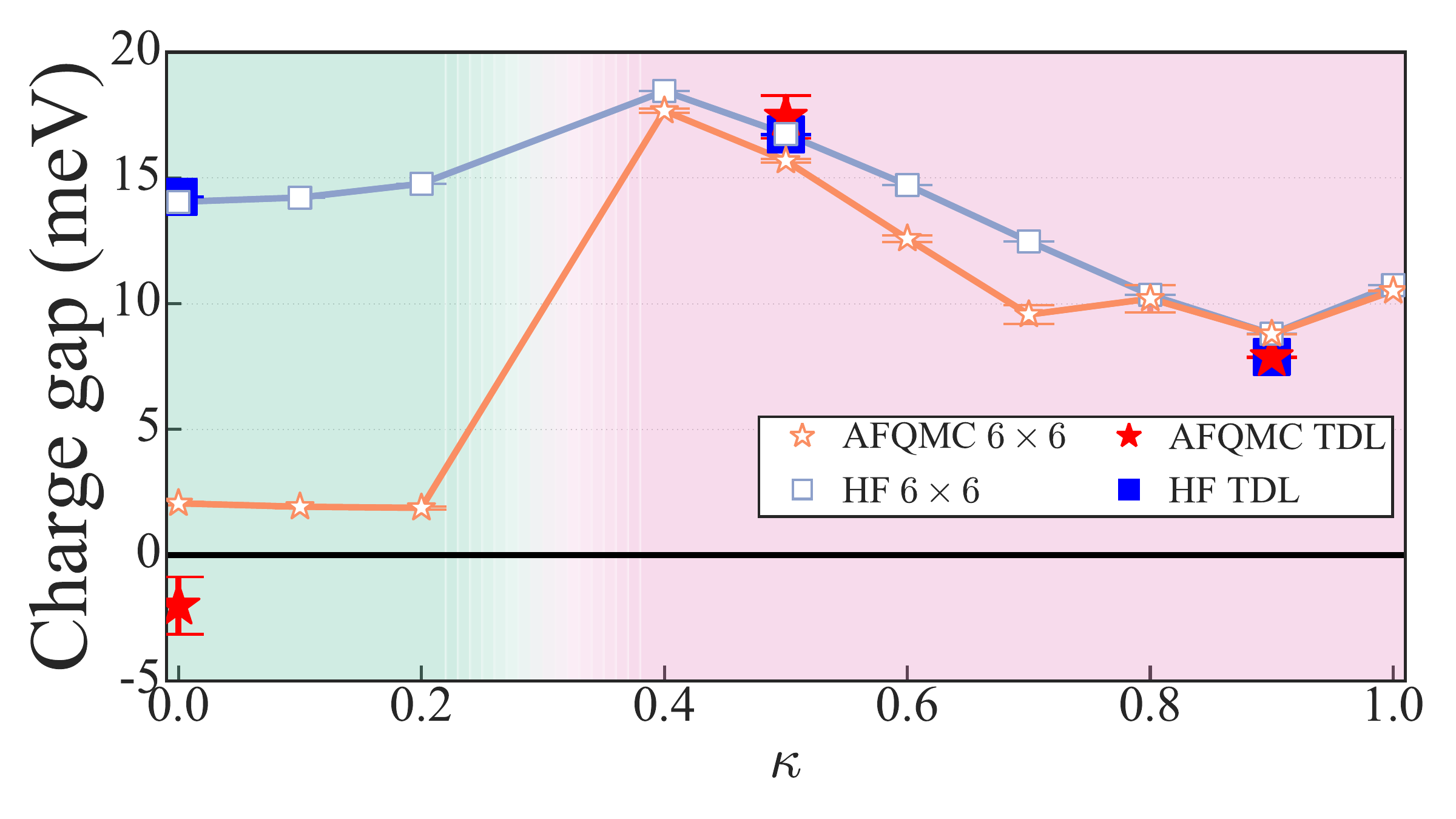} 
\includegraphics[scale = 0.2]{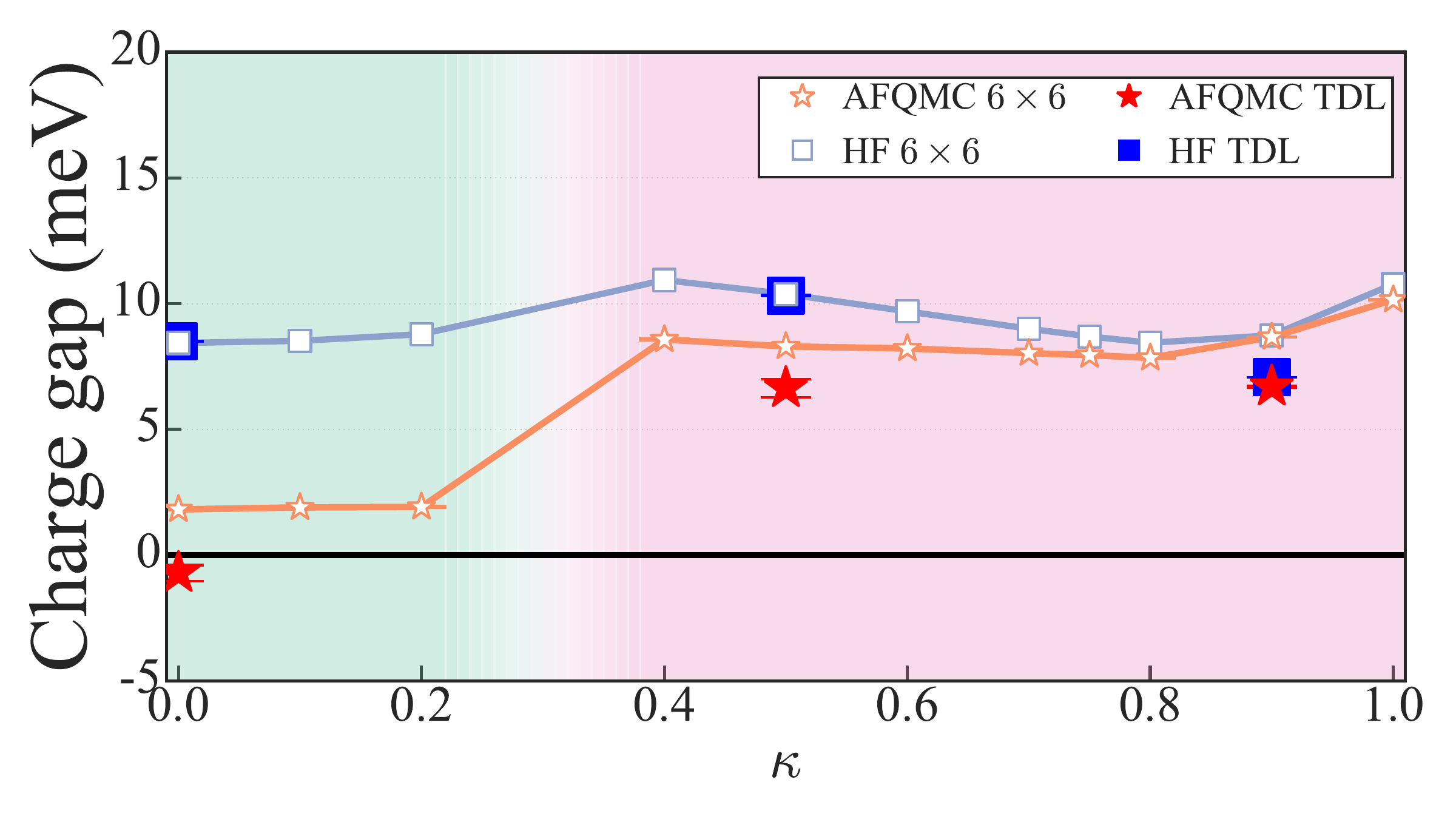} 
\caption{
Charge gaps  versus $\kappa$ 
at $n_f=4$ (top panel)
and $n_f=2$ (bottom panel). 
Similar to Fig.~\ref{Fig.CorrelationEnergy_CN+HF}, 
a $6 \times 6$ lattice is used in the 
calculations, but spot checks are 
performed at $\kappa=0.0$,$0.5$, and $0.9$ by extrapolating through $6 \times 6$, $9 \times 9$, and $12 \times 12$ lattices for the TDL. 
Shaded pink and green backgrounds separating the semi-metallic and insulating phases
are a rough guide to the eye. 
} 
\label{Fig.ChargeGap} 
\end{figure}

We next examine the ground state properties in some detail.
We compute the charge gap of the system via the add/remove energy:
\begin{equation}
\Delta_c = \frac{E(N+1)+E(N-1)-2\,E(N)\,,}{2}
\end{equation}
where $N$ is the total number of electrons in the simulation cell at the 
targeted filling fraction, and $E(N)$
is the total ground-state energy. 
The results are shown in Fig.~\ref{Fig.ChargeGap} as a function of $\kappa$ for both $n_f=4$ and 
$n_f=2$.
At $\kappa \approx 0.9$, AFQMC finds the same finite charge gaps as HF at both fillings, which is consistent with the picture from Fig.~\ref{Fig.CorrelationEnergy_CN+HF} and
confirms 
the K-IVC ground states in this regime. As $\kappa$ decreases, we observe a quantitative difference between the two filling fractions, 
with the charge gap increasing for $n_f=4$ and 
remaining essentially constant for $n_f=2$, until $\kappa$ reaches an intermediate value
around $0.4$, when a sharp 
drop in $\Delta_c$ is seen at 
both fillings. As $\kappa$ further decreases, the AFQMC charge gap is much reduced compared to 
the HF gap. A more careful finite-size scaling study, as illustrated for $\kappa=0$, shows that $\Delta_c$ from AFQMC becomes vanishingly small at the TDL, while the gap from from HF remains very large. 
Thus, contrary to the HF prediction,
the many-body state is metallic at small 
$\kappa$ at both fillings, with a metal-insulator transition taking place roughly around 
$\kappa \sim 0.3$.

\begin{figure}[htbp] 
\includegraphics[scale = 0.35]{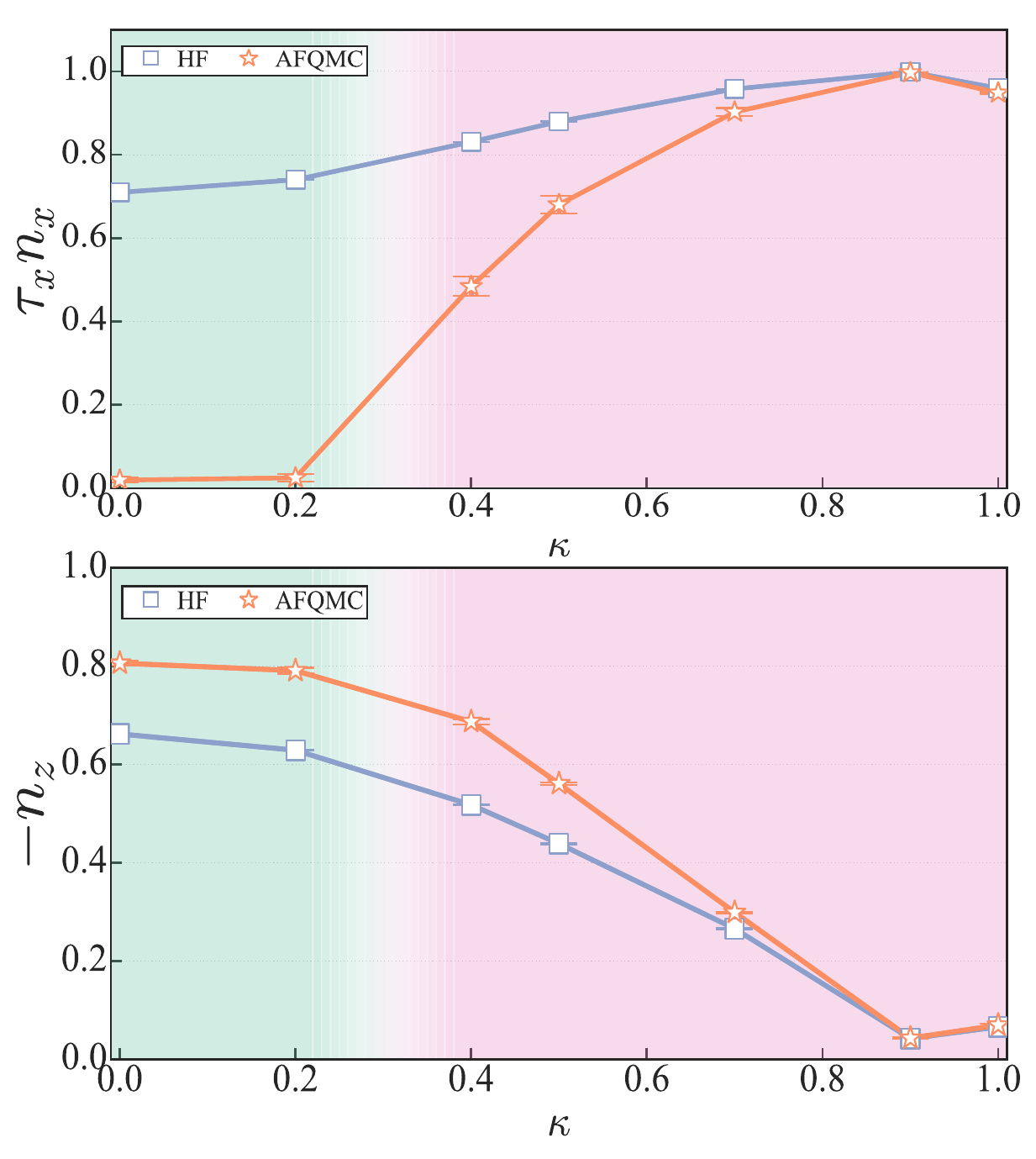} 
\caption{
K-IVC order parameter (top)
and band occupancy (bottom)
versus $\kappa$ 
at $n_f=2$ filling. 
Results are obtained with a $6\times 6$ lattice. 
Shaded pink and green backgrounds separate the metallic/semi-metallic and insulating phases as in Fig.~\ref{Fig.ChargeGap}.
}
\label{Fig.phaseTransition_HF}
\end{figure}

To better characterize the ground-state phases,
we also compute the K-IVC order parameter and band occupancy. 
Expectation values of operators that commute with the Hamiltonian, such 
as the ground-state energies and gaps discussed above, can be 
computed by the so-called mixed estimator
\cite{AFQMC_Zhang-Krakauer-2003-PRL} in AFQMC. Other expectations require 
additional 
technical steps to compute, 
for example by back-propagation
\cite{QMC_Zhang_Constrained_1997,AFQMC_PathRestoration} or via linear response \cite{Stripe_HaoXu, Absence_of_Superconductivity_MingPu}. We take the latter approach here, which requires separate calculations for different 
expectations but tends to be more accurate. More specifically,
we calculate the expectation value of 
an operator $\hat O$ with respect to the ground state from AFQMC,
$\langle \hat{O} \rangle$, via Hellman-Feynman theorem
\begin{equation}
\langle \hat{O} \rangle  = \underset{\lambda \to 0}{\lim}\frac{\mathrm{d} \langle \hat H_{IBM} +\lambda \hat{O} \rangle }{\mathrm{d} \lambda},
\end{equation}
which is then evaluated by finite difference
with 
several ground-state energy calculations at small $\lambda$. 
Further details are given in App.~\ref{sec:APPENDIX_order_parameter}.

In Fig.~\ref{Fig.phaseTransition_HF}, we show the computed K-IVC order parameter $\langle \tau_x\,n_x\rangle$
and band occupancy $\langle -n_z\rangle$
for different $\kappa$ at $n_f=2$ filling. 
The subscripts of $n_*$ and $\tau_*$ denote the related Pauli operator on bands and valleys (i.e., $\tau_x n_x=\frac{1}{L^2}\sum_{\textbf{k},s}c^\dagger_{\textbf{k},i'=(-n,-\tau,s)}c_{\textbf{k}, i=(n,\tau,s)}$, and $ n_z= - n \frac{1}{L^2}c^\dagger_{\textbf{k}, i}c_{\textbf{k}, i}$ with $\tau=1$ for $+$ valley and $\tau=-1$ for $-$ valley, $n=1$ for top band and $n=-1$ for bottom band.) 
The K-IVC order parameter 
is used to identify the strength of K-IVC order  (see App.~\ref{sec:APPENDIX_IBM_symmetry}) and the band occupancy 
shows the difference in electron distributions between the top and bottom bands. 
At $\kappa=0.9$, a strong signal of K-IVC order is seen, and electrons are almost evenly distributed over the top and bottom bands. These observations again confirm a K-IVC ground state. 
As $\kappa$ decreases, we find a decay of the K-IVC order in both the HF and AFQMC results. However, the K-IVC order in AFQMC decays much faster, and vanishes at $\kappa \sim 0.2$, while in HF it stays at a significant finite value. In the bottom panel 
the band occupancy increases as $\kappa$ decreases. A large $\langle -n_z\rangle $ implies a higher occupancy  of the bottom bands.
The computed value from AFQMC is always larger than from HF, indicating that the electrons tend to stay in the bottom bands more.
Since electrons forming K-IVC pairs tend to be evenly distributed over the top/bottom band, the results on 
$\langle -n_z\rangle $ suggest a decay of  K-IVC order with decreasing $\kappa$, consistent with $\langle \tau_x n_x\rangle$.
By symmetry \cite{Symmetry_of_TBG}, the above analysis and observations can be directly applied to $n_f=4$ filling.

We also examined the stability of a fully spin-polarized ground state for $n_f=2$ and non-spin-polarized ground state for $n_f=4$ by computing the energy cost for flipping spins. These results are included in App.~\ref{sec:APPENDIX_Low_lying_states}. 

\subsection{\label{sec:Q}Breaking of C$_{2}$T symmetry at $n_f=1$}

\begin{figure}[htbp]
\includegraphics[scale = 0.2]{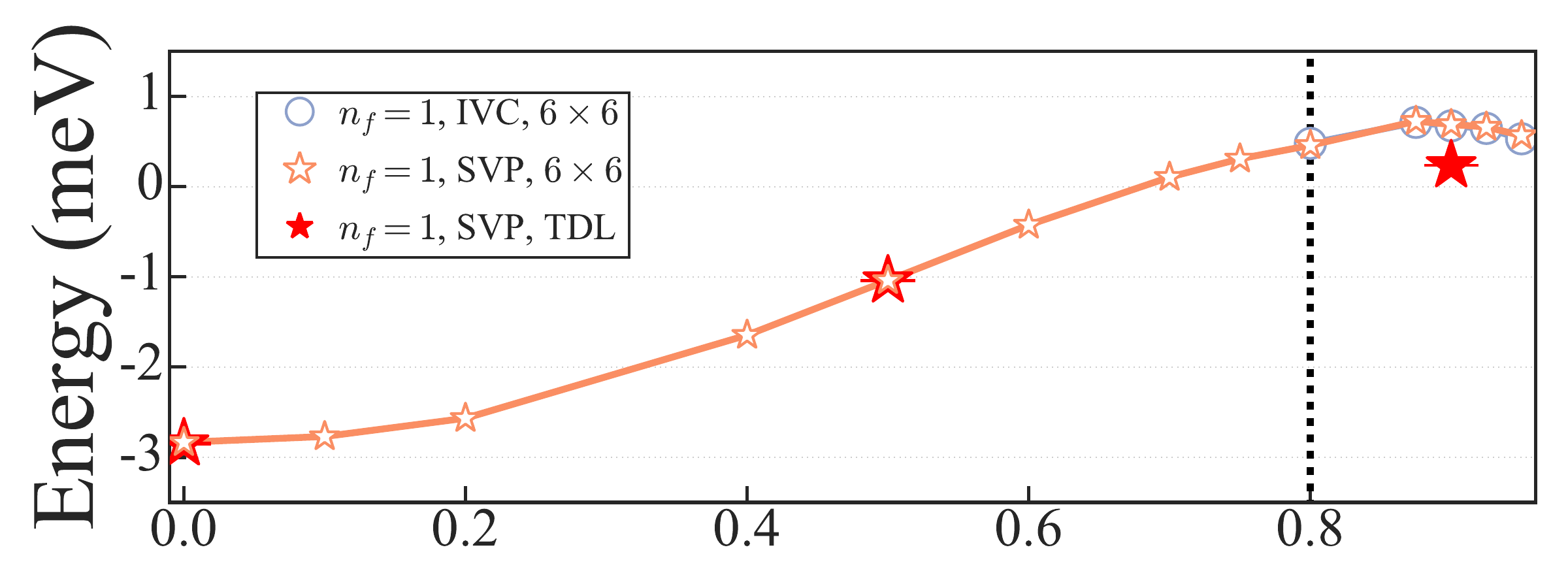} 
\includegraphics[scale = 0.2]{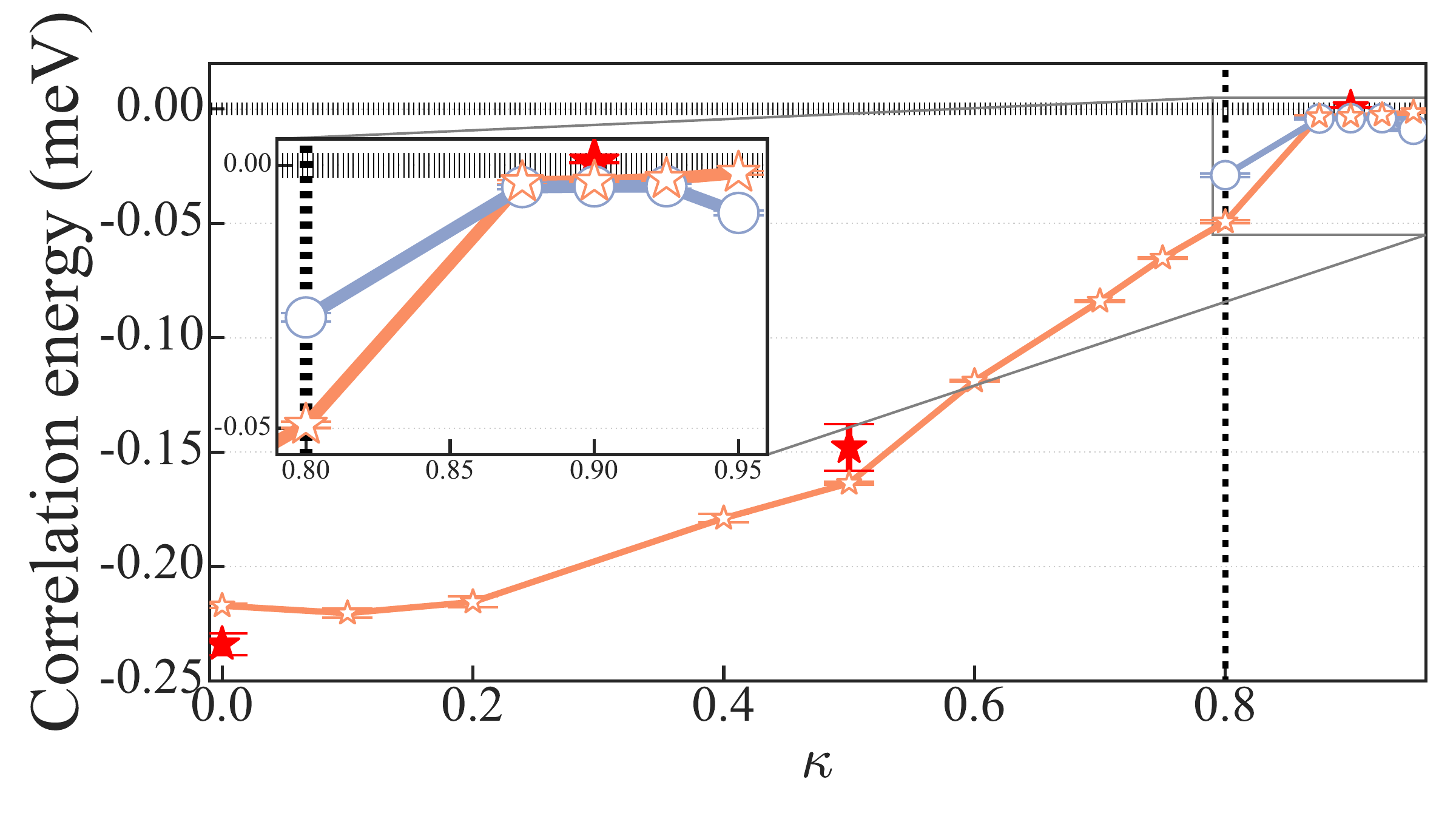} 
\caption{ 
Ground-state energies and correlation energies 
for spin valley polarized (SVP) state and inter-valley coherent (IVC) state, 
at $n_f=1$. The top panel shows the HF energy per electron, 
$E_{\rm HF}$, as a function of $\kappa$, 
for the two types of states considered. 
The bottom panel shows the corresponding
correlation energies, $E_c$. 
Most results are obtained from a $6\times 6$ 
lattice in momentum space, but at three values of $\kappa$ results are shown for the SVP state at the TDL, using the same extrapolation procedure as in Fig.~2.
Beyond $\kappa=0.8$ (indicated by the 
vertical black dashed lines) the SVP and IVC states are degenerate within HF, while small energy differences are seen in AFQMC.
However when extrapolated to the TDL at $\kappa=0.85$, the energy difference in AFQMC is zero within statistical error bar. 
}
\label{Fig.CorrelationEnergy_Q}
\end{figure}

At $n_f=1$, we find that the ground state can be reasonably approximated by HF through a comparison between HF and AFQMC results in both correlation energy and charge gap. This is consistent with results 
from a previous quantum chemistry study \cite{QC_LinLin_faulstich2022interacting}.
Based on HF, we observed a transition along $\kappa$ from the C$_{2}$T-symmetry-breaking correlated insulating state (i.e., Chern polarized QAH state \cite{Symmetry_of_TBG}) to a C$_{2}$T-symmetry-restored semi-metallic state and an emerged spin-polarized 
inter-valley-correlated 
semi-metallic state. 

In Fig.~\ref{Fig.CorrelationEnergy_Q}, we show HF energy in the top panel and the AFQMC correction over HF in the bottom panel for different $\kappa$ at $n_f=1$. Our HF study found a spin-valley-polarized (SVP) state for all $\kappa$, and a spin-polarized inter-valley coherent (IVC) state at $\kappa \geq 0.8$. 
In the bottom panel, the same correlation energy behaviors are seen as observed for $n_f=4,2$, where many-body effects are found to be strong at the chiral limit (i.e., $\kappa=0$) and 
become vanishingly small at large $\kappa$. 
However, the correlation energy for $n_f=1$ is much smaller than in $n_f=4,2$, and AFQMC only 
leads to a quantitative correction over HF.

\begin{figure}[htbp]
\includegraphics[scale = 0.2]{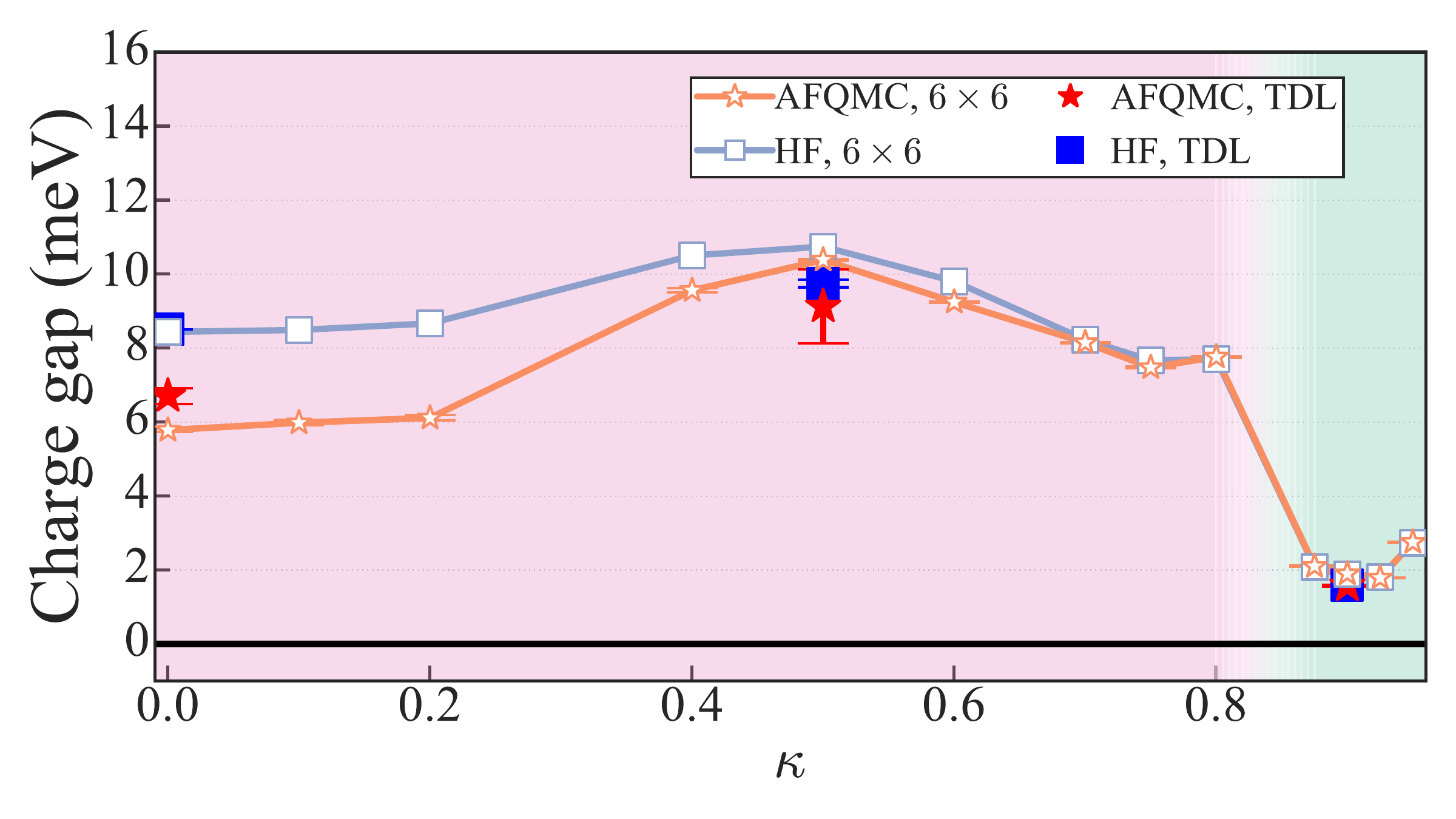}  
\caption{Charge gap for $n_f=1$ filling versus $\kappa$. Charge gaps are calculated by measuring the energy required to add or remove
an electron from the system. AFQMC proceeds with the corresponding HF state as the trial wave function, whose HF charge gap is illustrated in the figure. The thermodynamic limits (TDL) are extrapolated through 6×6, 9×9, and 12×12 lattice calculations. Shaded pink and green backgrounds separate the metallic/semi-metallic and insulating phases.
}
\label{Fig.charge_gap_Q}
\end{figure}

In Fig.~\ref{Fig.charge_gap_Q}, we show the charge gap for different $\kappa$. 
A finite charge gap is seen for $\kappa < 0.8$ at the thermodynamic limit, which suggests an 
insulating ground state. A sudden drop of the charge gap is observed around  $\kappa = 0.8$, indicating a transition to semi-metallic states. This behavior, however, is quite different from what is observed at $n_f=4,2$, where mean-field calculations do not capture the metallic states that emerge.

\begin{figure}[htbp]
\includegraphics[scale = 0.2]{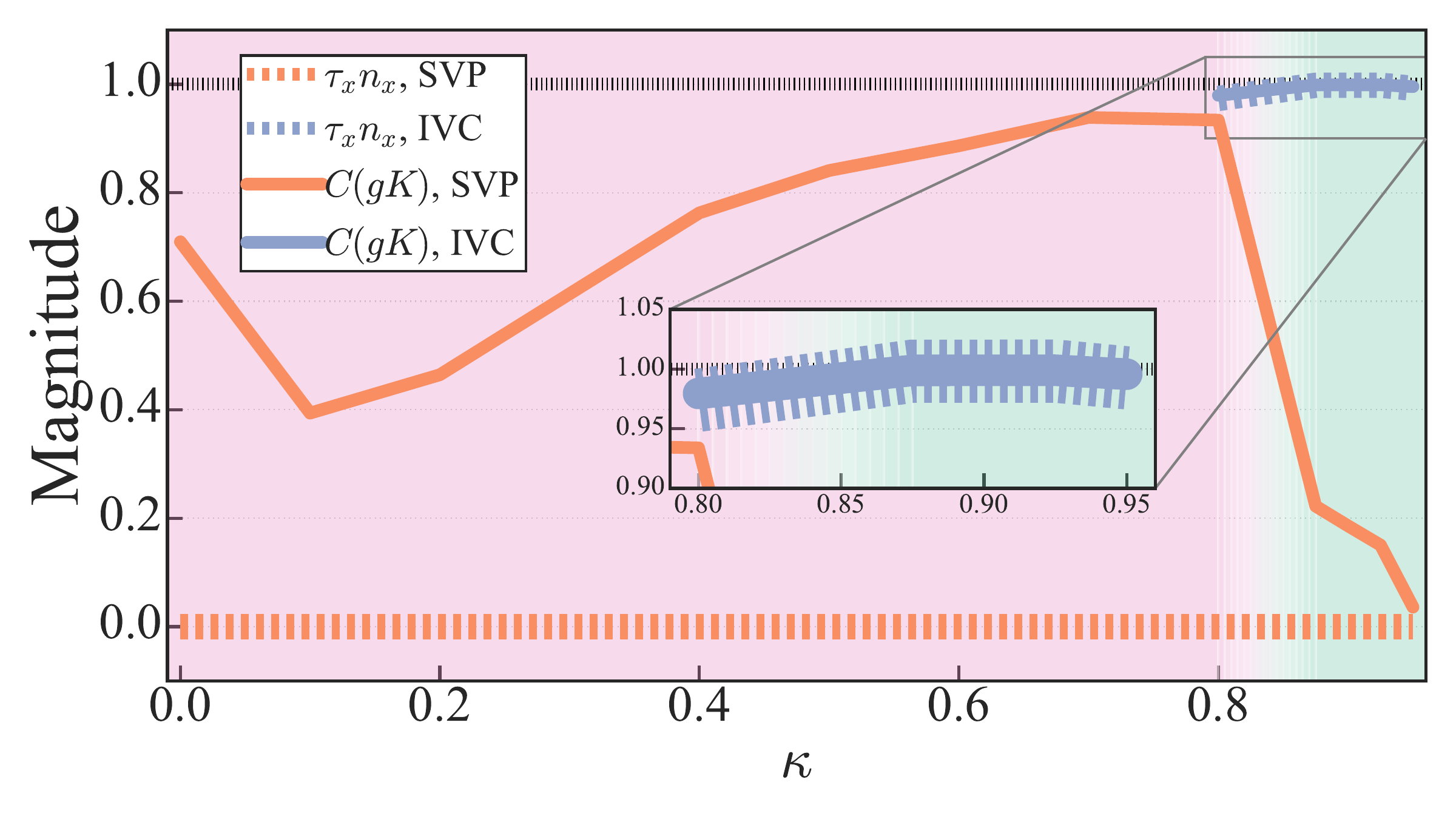}
\caption{K-IVC order parameter $ \tau_x n_x $ and C$_{2}$T-symmetry-breaking order parameters $C(gK)$ at $n_f=1$ for HF ground states with spin-valley-polarized (SVP) or Inter-valley coherent (IVC) symmetry breaking. In the degenerate region (i.e., between $\kappa=0.8$ and $\kappa=0.95$), HF states are determined by imposing the spin-polarized/inter-valley-coherent initial density, respectively. Shaded green and pink backgrounds separate the insulating and metallic/semi-metallic phases as in Fig.~\ref{Fig.charge_gap_Q}.
}
\label{Fig.phaseTransition_Q}
\end{figure}

In Fig.~\ref{Fig.phaseTransition_Q}, we show that the above insulating-semi-metallic transition 
is accompanied by a restoration of C$_{2}$T-symmetry and an emerge of IVC state. Since our energy and charge gap results show only minor differences between AFQMC and HF here, 
we consider the HF state as a qualitatively good approximation for the ground state, and 
we use HF to study the behavior of IVC and C$_{2}$T-symmetry by focusing on the K-IVC order parameter $\langle \tau_x n_x\rangle$ and the C$_{2}$T-symmetry-breaking order parameter $C(gK)$ proposed in \cite{QC_LinLin_faulstich2022interacting}, following identical notation
($g$ is unitary symmetry and $K$ is complex conjugation). $C(gK)$ vanishes when C$_{2}$T symmetry is completely restored. 
At small $\kappa$, the ground state is well described by a spin-valley polarized QAH state where C$_{2}$T-symmetry is significantly broken, and no K-IVC order is observed. As $\kappa$ increases, a degeneracy of the SVP state and spin-polarized IVC state is seen. For the SVP state, a sudden drop of $C(gK)$ indicates a restoration of C$_{2}$T-symmetry. For the spin-polarized IVC state, we find a significant K-IVC order parameter with significant C$_{2}$T symmetry breaking. Remarkably, $\tau_x n_x\sim 1$ suggests all electrons at $n_f=1$ form K-IVC pairs, but this is not a K-IVC state found in $n_f=4,2$ as there are not enough K-IVC pairs to fill entire valence bands.

\subsection{\label{sec:TQ} Mixture at $n_f=3$}
\begin{figure}[htbp]
\includegraphics[scale = 0.2]{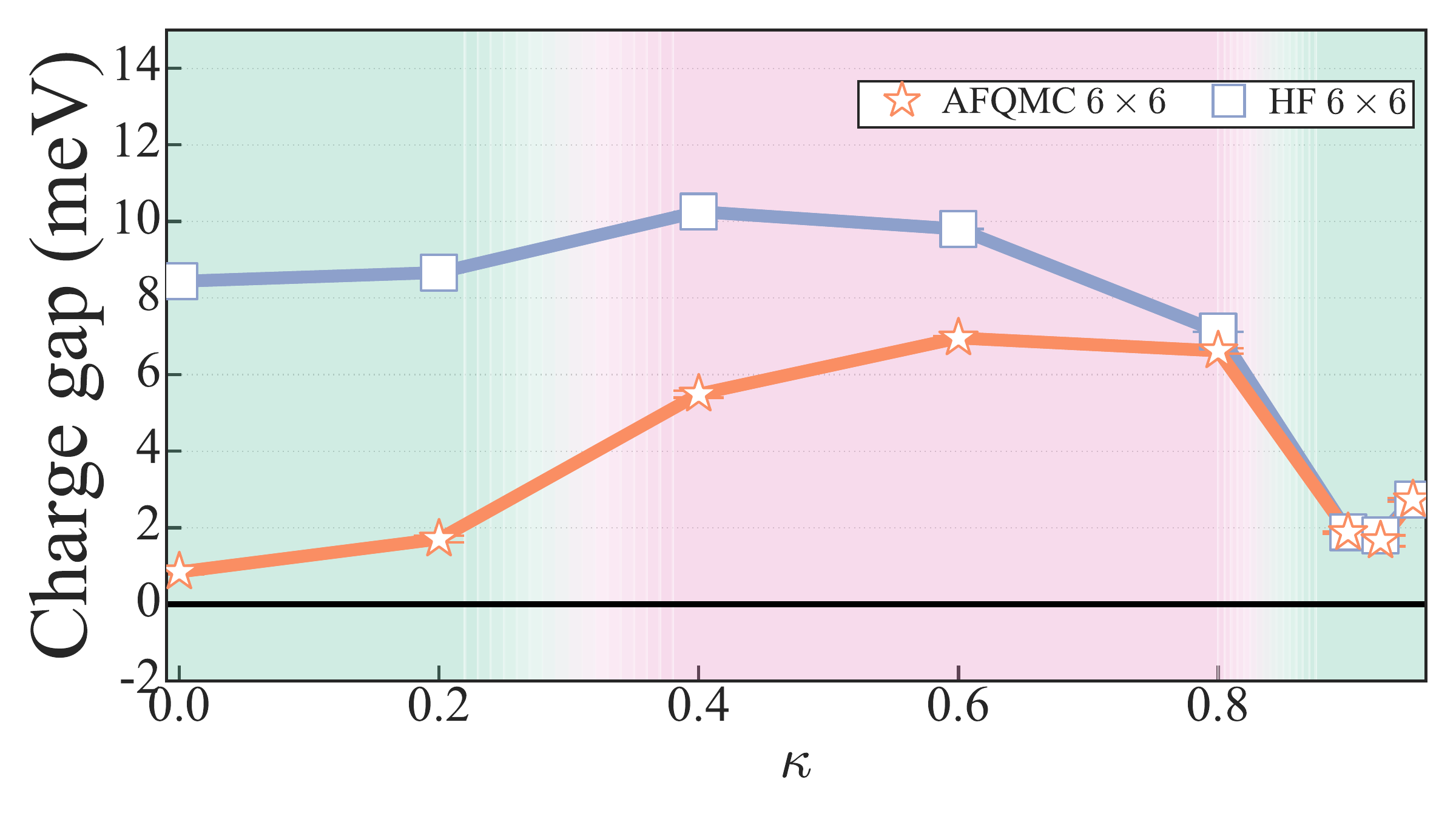}  
\caption{Charge gap for $n_f=3$ filling versus $\kappa$. 
AFQMC calculations use the corresponding HF state as the trial wave function, whose charge gap is also shown in the figure. Shaded pink and green backgrounds separate the metallic/semi-metallic and insulating phases.
}
\label{Fig.charge_gap_TQ}
\end{figure}

At $n_f = 3$ filling, we observed a ``phase diagram'' as a function of $\kappa$ that shows  features from both $n_f = 2$ and $n_f = 1$. In Fig.~\ref{Fig.charge_gap_TQ}, we show the charge gap for different $\kappa$. For the HF study, we observed a K-IVC state polarized at one spin flavor and a C$_{2}$T-symmetry-breaking state polarized at the other spin and one valley, which agrees with previous results \cite{HF_PRB2020}. For the AFQMC study, we also observed a combination of two states in the many-body sense. We observed a metallic state for small $\kappa$, and a transition to an insulating state as $\kappa$ is increased. As $\kappa \sim 0.8$ is approached, the drop in the charge gap is observed, which suggests a semi-metallic state. This behavior can be understood as a result of the direct combination between the state at $n_f=2$ and $n_f=1$.

\section{\label{sec:summary} Summary and discussion}
In this work, we investigated a model for twisted bilayer graphene that combines the BM model with realistic electron-electron Coulomb interactions. To minimize the difference between real materials and the model, we consider a mean-field correction to remove the double counting of Coulomb interactions and tune the ratio $\kappa$ to include the effects of lattice relaxation. We 
employ HF and AFQMC to study this 
model at three typical integer fillings,
in order to probe and quantify the effect of electron correlation.

We implemented state-of-the-art phaseless AFQMC to enable 
an accurate and systematic study of the ground state of the IBM model for a range of parameters and fillings.
By controlling the sign problem with the phaseless constraint, we perform calculations in the ground state and
in lattices as large as $12\times 12$ to reach the thermodynamic limit. Numerical Hartree Fork is first carried out to produce mean-field candidates of the ground states. Based on symmetry breaking in perturbation theories, these mean-filled states are correlated insulating states with either K-IVC or C2T symmetry breaking. We then use the 
HF states as trial wave functions for AFQMC and determine the actual ground states. 
The work presented here paves the way for systematic studies of realistic models of MATBG.
The same AFQMC technology we have set up can be directly applied to reach beyond the integer fillings studied here.
It can also be adopted to study the effect of Kekulé phonons via a non-retarded two-body attraction \cite{TBG_Phonon_1, TBG_Phonon_2}.

The outcome of these calculations is summarized as ground state phase diagrams for the IBM model for different integer fillings as a function of the tunneling ratio $\kappa$. 
Our results at charge neutrality are consistent with those from 
previous sign-problem-free QMC studies which focused on limited parameters \cite{DQMC_2022_Johannes, DQMC_2021_CPL, DQMC_Huang_2024_TBG_Qfilling}.
At charge neutrality and half-filling, we observe an insulator-metal transition with a loss of spin gap and K-IVC long-range order.  The results suggest that this transition is the result of many-body effects driven by competition between topological band and Coulomb interaction. At quarter-filling, we observe an insulator-metal transition that comes with a restoration of C$_2$T symmetry and an emerged inter-valley-correlated state. Our results indicate that the transition can be qualitatively described by HF. We observe the restoration of C$_2$T symmetry confirming conclusions of a previous study which did not include valley degree of freedom \cite{QC_LinLin_faulstich2022interacting}. An extra spin-polarized IVC state is seen in our study.
We also investigate the ground state phase diagram at three-quarter-filling. As a function of $\kappa$ it shows both features from both charge neutrality/half-filling and quarter-filling.

We find that the ground state of the IBM model is very sensitive to the corrections of the bare band structure.
This has also been seen in other many-body studies \cite{QC_LinLin_faulstich2022interacting, DQMC_2022_Johannes, DMRG_2021_Strain}.
Though conclusions from 
many-body studies, including ours, 
appear to yield qualitatively consistent results, the detailed model in each work involves different mean-field corrections, which lead to quantitative discrepancies. 
We believe it would be very valuable to have  
a single model to study. Agreement between different many-body calculations would then help ``converge'' 
a set of reliable solutions to the model. 
Discrepancies 
between different methods would help 
identify areas of improvement needed and spur method development and synergistic studies \cite{Hubbard_benchmark_2015,Mario_hydrogen_chain}. 

For example, recent experiments \cite{TBG_Quantum_textures} suggest the ground state of twisted bilayer graphene is in a time-reversal symmetric inter-valley coherent (T-IVC) phase, while a small strain can lead to incommensurate Kekulé spiral (IKS) order. Are these states captured by the IBM model but not by the detailed Hamiltonian we (and others) have studied? Or are there limitations in our computations? 
The aforementioned unique model and converged solution would allow more direct comparisons with experimental observations, which can then identify model limitations, and improve our understanding of essential physics.

\section{\label{sec:level7} Acknowledgments }
We thank L.~Lin,
A.~H.~MacDonald, O.~Vafek
for helpful conversations. 
Z.X. acknowledges partial support from the U.S. Department of Energy (DOE) under grant DE-SC0001303. Z.X.~is also grateful for the support and hospitality of the Center for Computational Quantum Physics (CCQ), where this work was performed. Computing was carried out at the computational facilities at Flatiron Institute. The Flatiron Institute is a division of the Simons Foundation.

\bibliography{cite} 

\section{\label{sec:APPENDIX_IBM}IBM model}

\subsection{\label{sec:BM_model}BM model }

We recap key features of the BM model \cite{BM_model, DQMC_2022_Johannes}.
To establish a tight-binding description of TBG, we first consider electrons on two graphene layers with no inter-layer interactions (i.e., the distance between two layers $a_0 \longrightarrow \infty$). We use unit cell index $l$, sub-lattice $\sigma$, and layer $\mu$ to identify the position of a selected carbon atom. We use $c^\dagger_{l,m,s}$, where $m=(\sigma,\mu)$, to specify the operator 
that creates an electron with spin $s$ 
in the $2p_z$ orbital
$\phi(\boldsymbol{r}-\boldsymbol{R}_{l,m})$ around the carbon atom at $\boldsymbol{R}_{l,m}$.

The Bloch state
\begin{equation}
\begin{aligned}
\Phi_{\boldsymbol{p},m}(\boldsymbol{r})=e^{i\boldsymbol{p} \cdot \boldsymbol{r}}u_{\boldsymbol{p},m}(\boldsymbol{r})
\end{aligned}
\label{eq:Bloch_State}
\end{equation}
where $u_{\boldsymbol{p},m}(\boldsymbol{r})=u_{\boldsymbol{p},m}(\boldsymbol{r}+\boldsymbol{R}_{l,m})$ is a periodic function, under tight-binding approximation, can be written as:
$$\Phi_{\boldsymbol{p},m}(\boldsymbol{r})=\sum_l e^{i\boldsymbol{p}\cdot \boldsymbol{R}_{l,m}}\phi(\boldsymbol{r}-\boldsymbol{R}_{l,m}).$$
In this appendix, 
$c^\dagger_{\boldsymbol{p},m,s}$ indicates  creating
a spin $s$ electron with Bloch state $\Phi_{\boldsymbol{p},m}(\boldsymbol{r})$.

\begin{figure}[htbp]
\includegraphics[scale = 0.65]{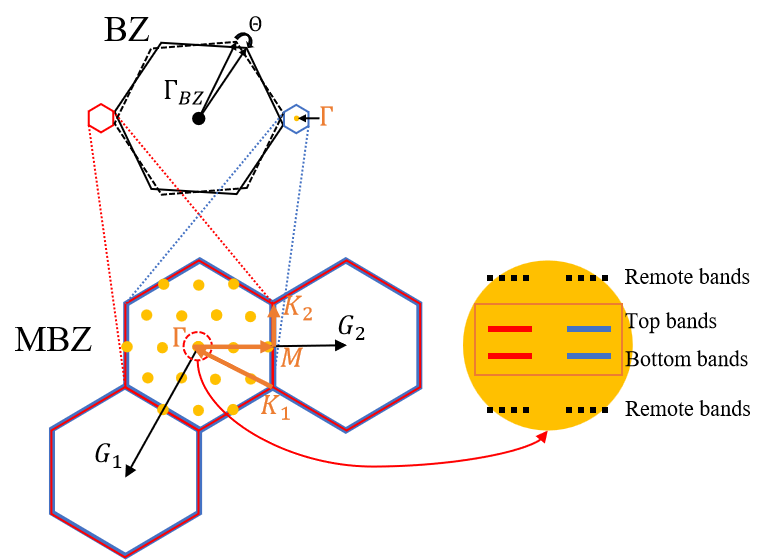}  
\caption{Two twisted graphene Brillouin zones (BZ) with twisted angle $\theta$ and the corresponding mini Brillouin zones (MBZ) for TBG moire 
lattice generated from the twist. The momentum space hexagonal lattice (i.e., yellow point) discretizes MBZ. $G_1$ and $G_2$ are primitive vectors of the MBZ. 
} 
\label{Fig.lattice} 
\end{figure}

As demonstrated in Fig.~\ref{Fig.lattice}, if two layers stack with a twisted angle $\theta$, the Brillouin zones (BZ) of one 
mono-layer of graphene rotate with angle $\theta$ relative 
to the other layer, and this rotation leads to a periodic moire pattern and a mini-BZ (MBZ). With the above notations, the tight binding Hamiltonian of bilayer graphene with no inter-layer hopping (for $\boldsymbol{p}$ measured around and from $K^\mu_{\pm}$ points) can be written as below:
\begin{equation}
\begin{aligned}
\hat{H}&=\sum_{\boldsymbol{p}}c^\dagger_{\boldsymbol{p}}T^{\textup{inter}}_{\boldsymbol{p},\boldsymbol{p}}c_{\boldsymbol{p}}\,\\
T^{\textup{intra}}_{\boldsymbol{p},\boldsymbol{p}}&=\hbar\upsilon _F(p_x\sigma_x\tau_z + p_y\sigma_y)e^{-i\theta \sigma_z \mu_z \tau_z /2}\,,
\label{Equ.no_inter_layer_interactions}
\end{aligned}
\end{equation}
where $\nu_F$ is Dirac velocity, $c^\dagger_{\boldsymbol{p}}=( ..., c^\dagger_{\boldsymbol{p},m,s} , ... )$ is a row vector of electron creation operators. $\boldsymbol{p}=(p_x,p_y)$ run over the 1st BZ of graphene at layer $\mu$ and the choice of the $K^\mu_{\pm}$ point is denoted by valley $\tau$. $\tau_\star$, $\sigma_\star$, and $\mu_\star$ are the Pauli matrices for the related flavors.

By incorporating 
inter-layer hopping into the tight-binding Hamiltonian and a $U_g=e^{-i\theta \sigma_z \mu_z \tau_z /4}$ basis transformation (i.e., $c_{\boldsymbol{p}} \rightarrow U_g c_{\boldsymbol{p}}$ ), we have:
\begin{equation}
\begin{aligned}
\hat{H}_{BM}&=\sum_{\boldsymbol{p}}T^{\textup{intra}}_{\boldsymbol{p},\boldsymbol{p}}c^\dagger_{\boldsymbol{p}}c_{\boldsymbol{p}} + T^{\textup{inter}}_{\boldsymbol{p}',\boldsymbol{p}}c^\dagger_{\boldsymbol{p}'}c_{\boldsymbol{p}}\\
T^{\textup{intra}}_{\boldsymbol{p},\boldsymbol{p}}&=\hbar\upsilon _F(p_x\sigma_x\tau_z + p_y\sigma_y)\\
T^{\textup{inter}}_{\boldsymbol{p}',\boldsymbol{p}}&=\frac{1}{2}\sum_{l=0,1,2}\delta_{\boldsymbol{p}',\boldsymbol{p}-\tau_z \boldsymbol{q}_l}(\mu_x-i\mu_y)T_l + \textup{h.c.}\\
T_l&=\omega_0e^{-i\theta \sigma_z \mu_z \tau_z/2} + \omega_1e^{2\pi i l \sigma_z \tau_z/3} \sigma_x e^{-2\pi i l \sigma_z \tau_z/3}\,,
\label{Equ.BM}
\end{aligned}
\end{equation}
where $\theta = 1.05^{\circ}$ is the twist angle,  $\nu_F=2700*\frac{3a_0}{2\hbar}$ meV is the Dirac velocity, and $\omega_0$, $\omega_1$
describes lattice relaxation between A-A and A-B stacks, respectively. In this work, we take $\omega_1=105$meV and tune $\kappa=\frac{\omega_0}{\omega_1}$ to scan $\omega_0$.  Transition momentum $\boldsymbol{q}_l=R(2\pi l/3)\boldsymbol{q}_0$, where $R(\psi)=e^{i\psi \sigma_y}$ is a rotation matrix. $\boldsymbol{q}_0=\frac{8\pi \textup{sin}(\theta/2)}{3\sqrt{3}a_0}(0,-1)^T$ and $a_0=1.42 \AA$ is the intra-layer distance between carbon atoms.

Since the model $\hat{H}_{BM}$ \cite{BM_model} above are described in ``microscopic basis" (i.e., unit cell index $l$, sub-lattice $\sigma$ and layer $\mu$), we transform it to ``band basis" that simplifies the implementation of AFQMC. As demonstrated in Fig.~\ref{Fig.lattice}, we discretize MBZ at valley $\tau$ as $L\times L$ lattice, and each lattice point is labeled by $\boldsymbol{k}+\boldsymbol{G}=(k_1+m_1*G_1,k_2+m_2*G_2)$ with $\boldsymbol{k}$ running over the 1st MBZ and $G_1, G_2$ are the primitive vectors of MBZ. $\Gamma$ point of the hexagon is set to be the origin $\boldsymbol{k}=(0,0)$. We shift $\boldsymbol{p}$ with $\boldsymbol{p}+K^{\mu}_{\pm}=\boldsymbol{k}+\boldsymbol{G}+\Gamma$, then $\hat{H}_{BM}$ is block-diagonalized for $\boldsymbol{k}$, spin $s$, valley $\tau$ where only terms that connect $\boldsymbol{k}$ and $\boldsymbol{k}+\boldsymbol{G}$ are non-zero. In the numerical study, we truncate $\boldsymbol{G}$ with the closest $N_G$ vectors ($-4 \le m_1, m_2 \le 4$ in this work). Then we can use $\varepsilon _{\boldsymbol{k},i}$ and $|\varphi_{\boldsymbol{k},i} \rangle$ to specify eigenvalues and eigenstates of each band $n=0,...,N_G-1$ in this model:
\begin{equation}
\hat{H}_{BM}|\varphi_{\boldsymbol{k},i} \rangle=\varepsilon _{\boldsymbol{k},i}|\varphi_{\boldsymbol{k},i} \rangle
\end{equation}
which is in band basis labeled by $i=(s,\tau,n)$ and Bloch wave functions $|\varphi_{\boldsymbol{k},i} \rangle=\sum_{\boldsymbol{G},m}\alpha_{\boldsymbol{k}+\boldsymbol{G},i,m}|\Phi_{\boldsymbol{k}+\boldsymbol{G},m}\rangle$.

We limit this model to two flat bands closest to the Fermi level to allow a practical numerical study. We only consider one-body and two-body terms involving two flat bands $n=\pm$ where $\pm$ specify the top and bottom flat bands. When $|\varphi_{\boldsymbol{k},i} \rangle$ are degenerated, we fix the gauge with C$_{2}$T symmetry as used in previous DQMC work \cite{DQMC_2022_Johannes}. 

\subsection{\label{sec:IBM_model}Interacting BM model }

The two-body interaction is already 
included in the BM model at an independent-electron level through the original
ab initio density-functional computations.
To  include the screened Coulomb interaction, 
the IBM model in Eq.~(\ref{Equ.IBM})
also needs to include a background subtraction to correct this double-counting effect.
The Coulomb interaction $\hat{V}$ is limited in two flat bands, and the charge-density operators are shifted according to the chemical potential at $n_f$ filling:
\begin{equation}
\begin{aligned}
&\bar{\rho}_{\boldsymbol{q}} = \sum_{\boldsymbol{k},i,j} d^\dagger_{\boldsymbol{k}+\boldsymbol{q},i}\Lambda (\boldsymbol{k}+\boldsymbol{q},\boldsymbol{k})_{i,j} d_{\boldsymbol{k},j} - \frac{n_f}{8}\!\!\sum_{\boldsymbol{k}}\delta_{\boldsymbol{q},\boldsymbol{G}}\textup{tr}\{\Lambda (\boldsymbol{k}+\boldsymbol{q},\boldsymbol{k})\}\\
&\Lambda (\boldsymbol{k}+\boldsymbol{q},\boldsymbol{k})_{i,j}\equiv \sum_m\alpha^\star_{\boldsymbol{k}+\boldsymbol{q},i,m}\alpha_{\boldsymbol{k},j,m}
\end{aligned}
\end{equation}
where $d^\dagger_{\boldsymbol{k},i}$ create state $|\varphi_{\boldsymbol{k},i} \rangle=\sum_{\boldsymbol{G},m}\alpha_{\boldsymbol{k}+\boldsymbol{G},i,m}|\Phi_{\boldsymbol{k}+\boldsymbol{G},m}\rangle$ and momentum $\boldsymbol{k}$ that runs over MBZ. The transfer momentum $\boldsymbol{q}$ remains unrestricted but is truncated for numerical implementation which is $ (-3*G_1,-3*G_2) \le \boldsymbol{q}  \le (3*G_1,3*G_2)$ in this work. The potential
\begin{equation}
V_{\boldsymbol{q}}=\frac{e^2}{2\epsilon \epsilon _0}\frac{1}{A}\frac{\textup{tanh}(|\boldsymbol{q}|d)}{|\boldsymbol{q}|}
\end{equation}
is the Fourier-transformed two-gate screened Coulomb interaction. $A = N_M A_0$ is the total area of the system with $A_0=\sqrt{2}a^2_M/2$, $a_M=\sqrt{3}a_0/2 \textup{sin}\frac{\theta}{2}=13.4$ nm and $N_M$ is the number of moire unite cells of this system. We chose a distance $d=20$ nm between two layers and relative permittivity $\epsilon=10$ ($\epsilon_0$ is vacuum permittivity). We study the choice of relative permittivity in App.~\ref{sec:APPENDIX_parameter}.

The inclusion of the Coulomb interaction in $\hat{V}$ should be accompanied by a removal of the double-counting of Coulomb interacting in $\hat{H}_{BM}$, as discussed earlier. The ``bare" dispersion in TBG without Coulomb interaction involved is approximated with
\begin{equation}
\begin{aligned}
H_0&=\hat{H}_{BM} - [\hat{V}]_{\phi_{BM}} \\
&=\sum_{\boldsymbol{k},i}\varepsilon _{\boldsymbol{k},i}c^\dagger_{\boldsymbol{k},i}c_{\boldsymbol{k},i} - \sum_{\boldsymbol{k},i,j}[V]_{\boldsymbol{k},i;\boldsymbol{k},j}c^\dagger_{\boldsymbol{k},i}c_{\boldsymbol{k},j}.
\end{aligned}
\end{equation}
with a mean-field approximation $[V]$ of the Coulomb interaction. The choice of mean-field state to estimate the double counting correction $[V]_{\boldsymbol{k},i;\boldsymbol{k},j}$ can be varied. This paper uses the ground state of $H_{BM}$ at $n_f=4$ as our mean-field reference state $\phi_{BM}$. The discrepancies between bare dispersion $H_0$ and BM model $H_{BM}$ along $\kappa$ are presented in Fig.~\ref{Fig.H_bandstructure}.

In practice, we first write $\hat{H}_{BM}$ (Eq.~\ref{Equ.BM}) in ``microscopic basis" where basis are labeled with spin $s$, valley $\tau$, layer $\mu$ and sublattice $\sigma$. We then diagonalize $\hat{H}_{BM}$ for flat bands closest to the fermi surface and identify two bands through energy. For two flat bands that are degenerated (i.e., $\epsilon_{\boldsymbol{k}}=0$), we break the degeneracy with C$_{2}$T and ``small angle approximation" \cite{DQMC_2022_Johannes} by taking their eigenvectors as our specified bands. We take the set of bands and related spin, valley, as our band basis (labeled by $n$, $s$, $\tau$) where $H_{BM}$ is diagonalized.

\section{\label{sec:APPENDIX_IBM_symmetry}Symmetry in IBM model}
According to a previous study \cite{Symmetry_of_TBG, skyrmions_topology_of_TBG}, at the spinless limit, the ground states of IBM (without double counting correction) at chiral and flat band limit are single determinants: 
\begin{itemize}
    \item ( Chern number $\pm 2$) fill two bands in the same Chern sector $\longrightarrow$ quantum anomalous Hall states (QAH $\sigma_z\tau_z$). 
    \item ( Chern number $0$) fill as linear combinations of the two bands in both Chern sectors $\longrightarrow$ Valley Polarized states(VP $\tau_z$), Valley Hall states(VH $\sigma_z$), T-symmetric Inter-valley-coherent states (T-IVC $\sigma_x\tau_{x,y}$)and Kramers inter-valley-coherent states(K-IVC $\sigma_y\tau_{x,y}$).
\end{itemize}

The limit away from flat band can be considered perturbatively as an emergence of dispersion term:
$$
h(\boldsymbol{k})=h_0(\boldsymbol{k})\tau_z+ h_x(\boldsymbol{k})\sigma_x+h_y(\boldsymbol{k})\sigma_y\tau_z
$$ 
that tunnel between opposite Chern (sublattice) bands belonging to the same valley, and we have
\begin{itemize}
    \item At $\kappa=0$, QAH, VH, and K-IVC are exact ground states if there is no dispersion. If dispersion is non-zero, the perturbation of the dispersion indicates the candidates of ground state prefer QAH, VH, and K-IVC. 
    \item At $\kappa>0$, the combination of interaction terms (an extra sublattice exchange term should be considered when $\kappa>0$) and the perturbation of the dispersion prefer K-IVC.
\end{itemize}

HF \cite{Symmetry_of_TBG, HF_PRB2020, HF_PRR2021} can produce the above correlated insulating states respectively by imposing corresponding symmetry breaking in the initial density. If the dispersion term is small, the breaking of $U_V(1)$ valley charge conservation indicates 
K-IVC state and the unbroken of $U_V(1)$ valley charge conservation (no inter-valley terms) combined with C$_{2}$T symmetry helps us to distinguish VH and QAH (C$_{2}$T symmetry is broken) with VP and SP state (C$_{2}$T symmetry is preserved). However, for a non-zero dispersion term, especially when perturbation fails, the solution of HF has no need to be eigenstates of QAH, VH, K-IVC correlation (i.e., exactly VH, QAH, and K-IVC state at flat band limit). In this paper, the term ``K-IVC state" specifies HF ground state that have non-zero IVC, and ``C$_{2}$T-symmetry-breaking state" summarize all HF ground states that have no IVC and breaks C$_{2}$T symmetry (i.e., QAH and VH) as our study does not distinguish VH and QAH.

\section{\label{sec:APPENDIX_Low_lying_states}Low-lying states around chiral limit for $n_f=4,2$ fillings}

\subsection{Spin gaps}
\begin{figure}[htbp]
\includegraphics[scale = 0.185]{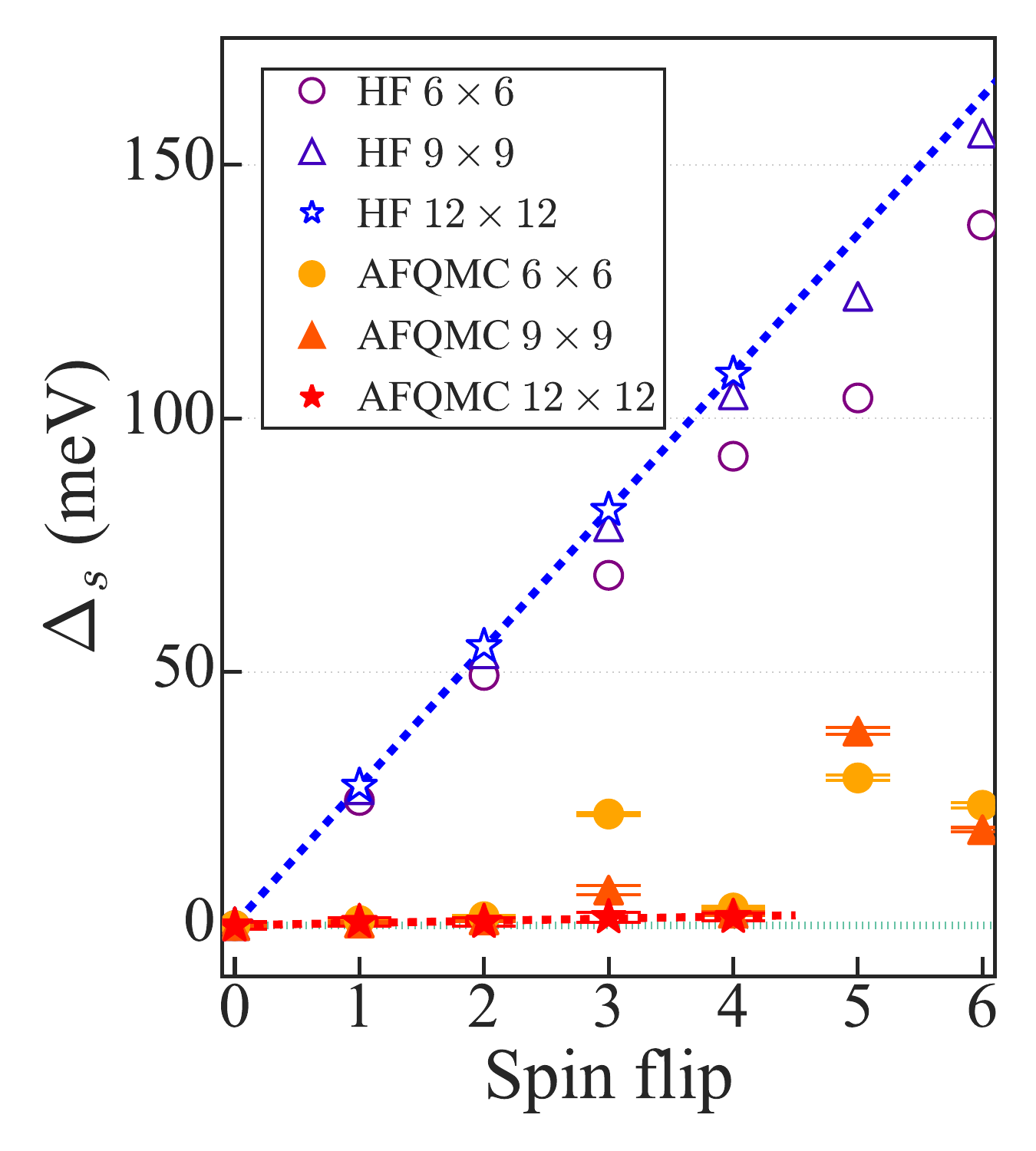}
\includegraphics[scale = 0.185]{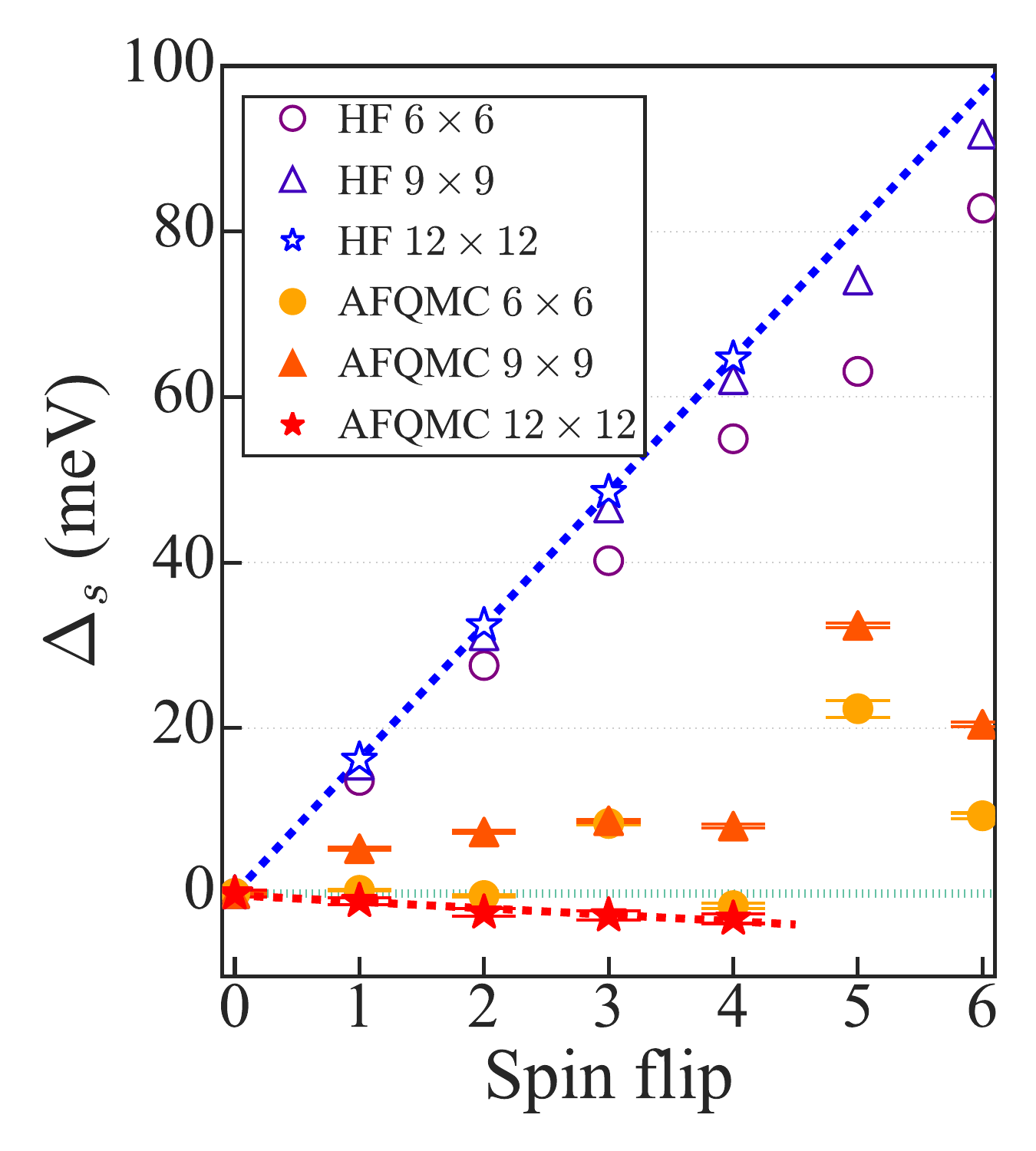}
\caption{
Spin gap at the  chiral limit ($\kappa=0.0$) for fillings $n_f=4$ (left panel) and  $n_f=2$ (right panel).
$\Delta_s$ is shown as a function of $\delta N$ for a small number of spin flips. 
States with zero spin-flip corresponds to the lowest HF energy.
Dashed lines connect results from the largest lattice sizes $12\times 12$.
}
\label{Fig.Energy_SpinFlip_CN+HF}
\end{figure}

We take the chiral limit for $n_f=4,2$ fillings.
To determine the spin polarization, we calculate the spin gaps 
\begin{equation}
\Delta_s = E(N_\uparrow-\delta N,N_\downarrow+\delta N) -
E(N_\uparrow,N_\downarrow)\,,
\end{equation}
where $\delta N=1,2, \cdots$.  
In Fig.~\ref{Fig.Energy_SpinFlip_CN+HF}, we show the 
spin gaps as a function of spin flips ($\delta N$) 
at $\kappa=0.0$ from 
both HF and AFQMC. 
The reference states, with zero spin-flip, are the states with the lowest HF energy, namely, the non-spin polarized ground state ($N_\uparrow=N_\downarrow=N/2$) at $n_f=4$ and fully spin-polarized state 
($N_\uparrow=N$, $N_\downarrow=0$) for $n_f=2$ fillings. 
The AFQMC calculations used the corresponding HF state as a trial wave function.   
We show results from several lattice sizes, up to $12\times 12$, to help gauge finite-size effects. Our data 
support a 
vanishing spin gap around zero spin-flip.

\subsection{Constraint robustness and further correction with constraint release}
Many studies \cite{Hubbard_benchmark_2015, simons_material_2020, Mario_hydrogen_chain} have shown that AFQMC can recover from a constraining trial wave function which is incorrect. 
However there has been no previous application of AFQMC to the MATBG system.
Here we illustrate how in the most ambiguous cases we can reduce the dependence on  which low-lying states as the trial wave function by constraint release.

\begin{figure}[htbp] 
\includegraphics[scale = 0.17]{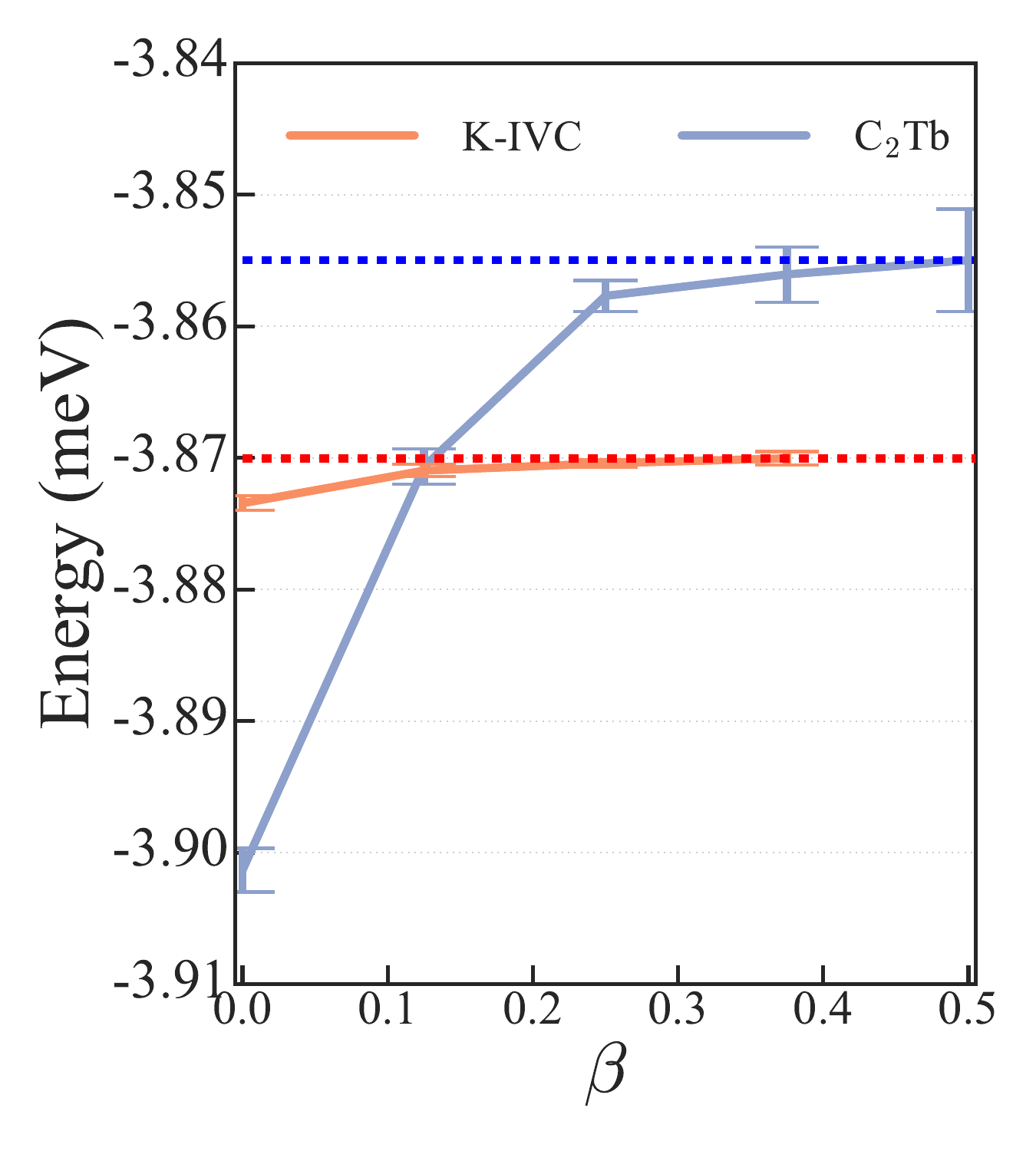} 
\includegraphics[scale = 0.17]{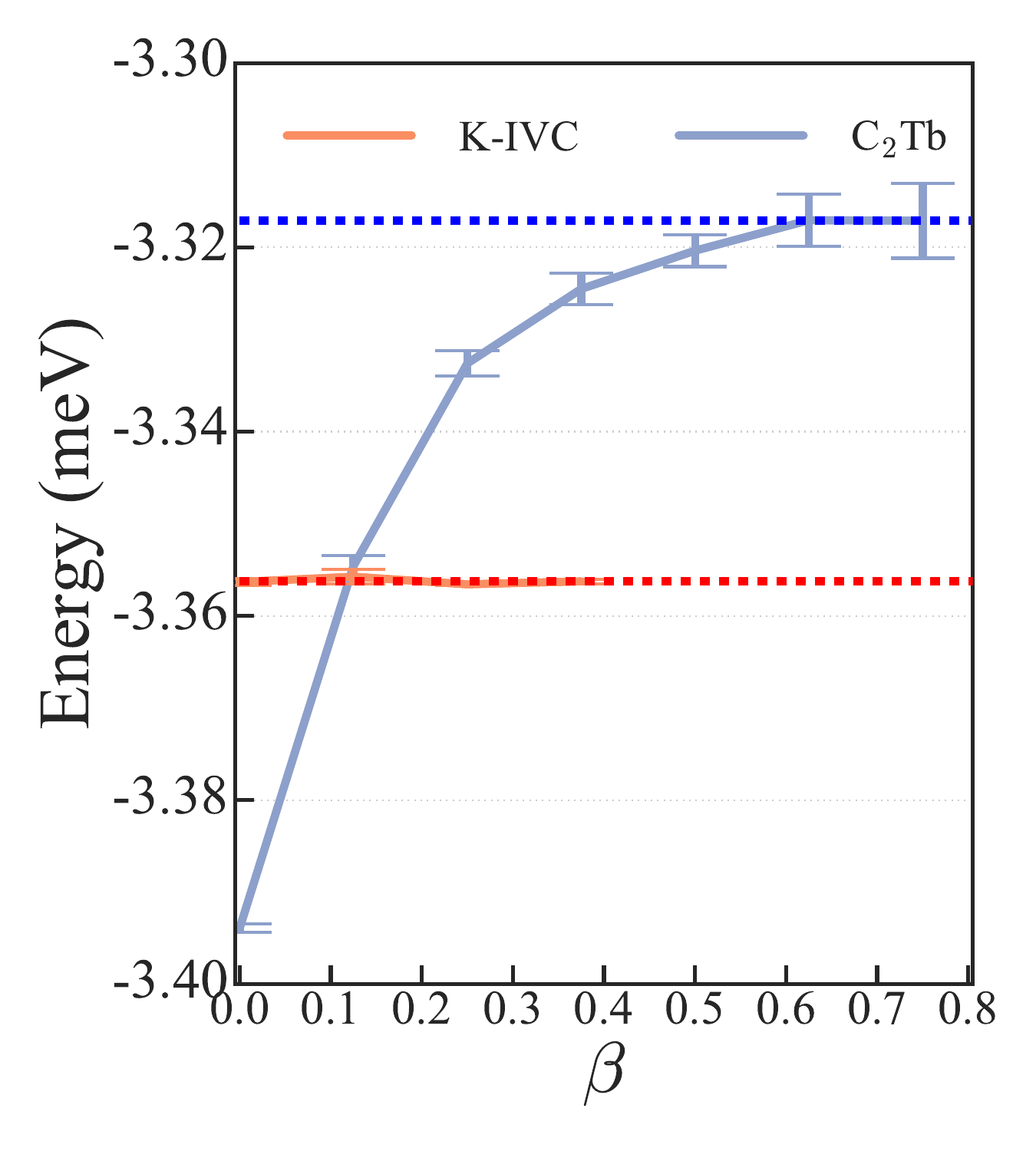}
\caption{
Quantifying and reducing the constrained path error by AFQMC with Metropolis release constraint \cite{zxiao_MRC_2023}.
Energy per electron vs. released imaginary time $\beta$ at $\kappa=0$ for charge neutrality $n_f=4$ (left) and half-filling $n_f=2$ (right). Energy from AFQMC with K-IVC state or C$_2$T-symmetry-breaking state as the trial wave function is identified with orange and blue lines, respectively. Dashed lines indicate the converged energy for the Metropolis release procedure. }
\label{Fig.Metro}
\end{figure}

To distinguish low-lying states that are closely competing (i.e., K-IVC state and C$_{2}$T-symmetry-breaking state, which are degenerated at the chiral limit in the HF study), we detect ground state preference between these two states by comparing the energy difference between AFQMC with corresponding trial wave functions. That is the AFQMC run with lower energy gives a better description of the ground state. To resolve such a small difference, AFQMC with Metropolis release constraint \cite{zxiao_MRC_2023} is performed to detect -- and eliminate -- any residual systematic bias from the constrained path approximation. 
AFQMC with Metropolis release can exponentially reduce the systematic bias by increasing released imaginary time $\beta$. In Fig.~\ref{Fig.Metro}, we show the detailed convergence in energy for $\kappa=0$. In Fig.~\ref{Fig.Energy_KIVC_C$_{2}$T_Kappa_CN+HF}, we summarise the comparison of converged energy in AFQMC with Metropolis release for different trial wave functions and different $\kappa$.  We show that AFQMC with K-IVC trial wave function has lower energy for both $n_f=4,2$ fillings. 
Note that the competition between the ground states in K-IVC and C$_2$T-symmetry-breaking manifolds is only relevant when $\kappa$ is small. The energy of ground states in K-IVC manifolds is much lower than the others for $\kappa \gg 0$. 

\begin{figure}[htbp] 
\includegraphics[scale = 0.17]{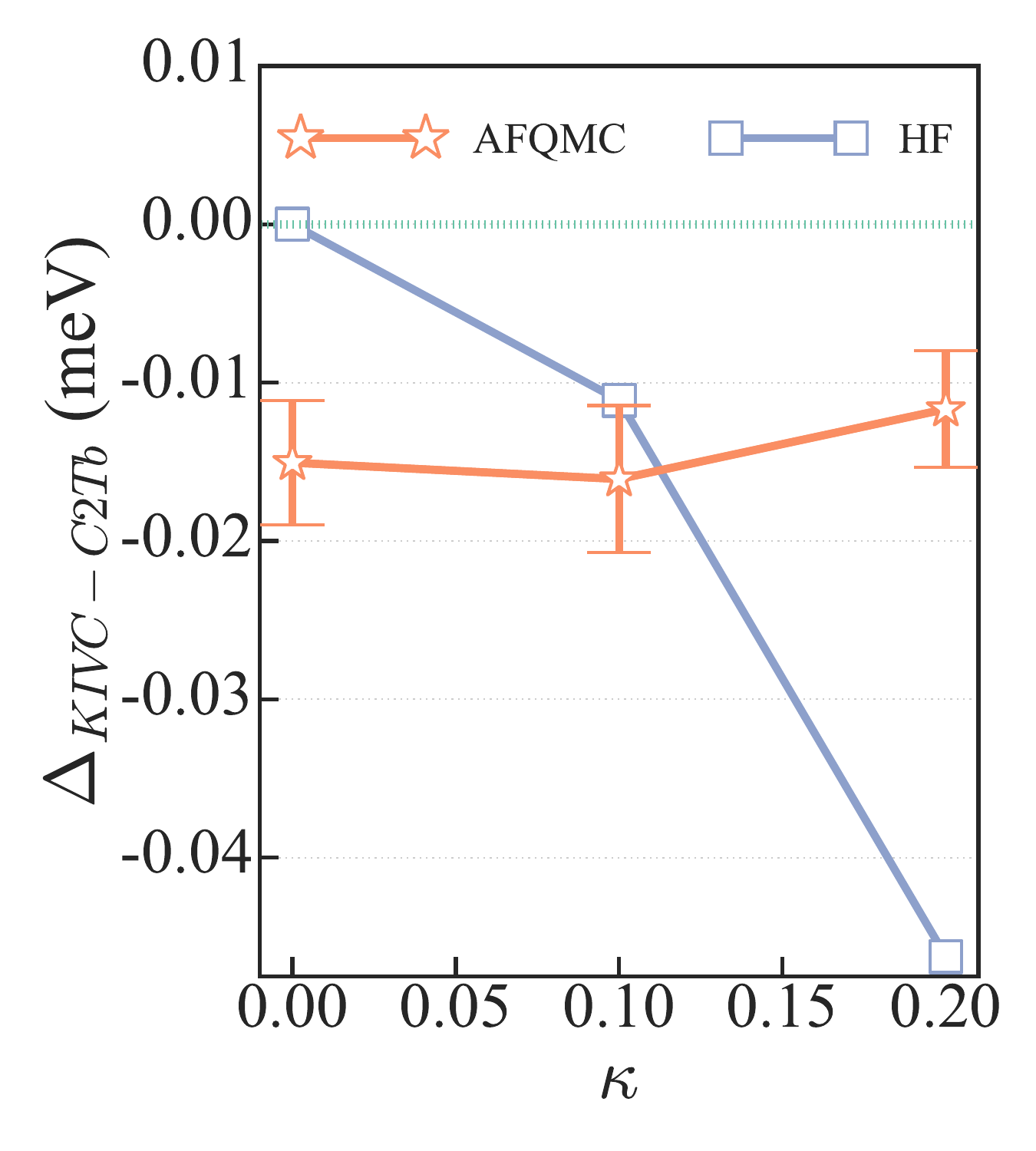} 
\includegraphics[scale = 0.17]{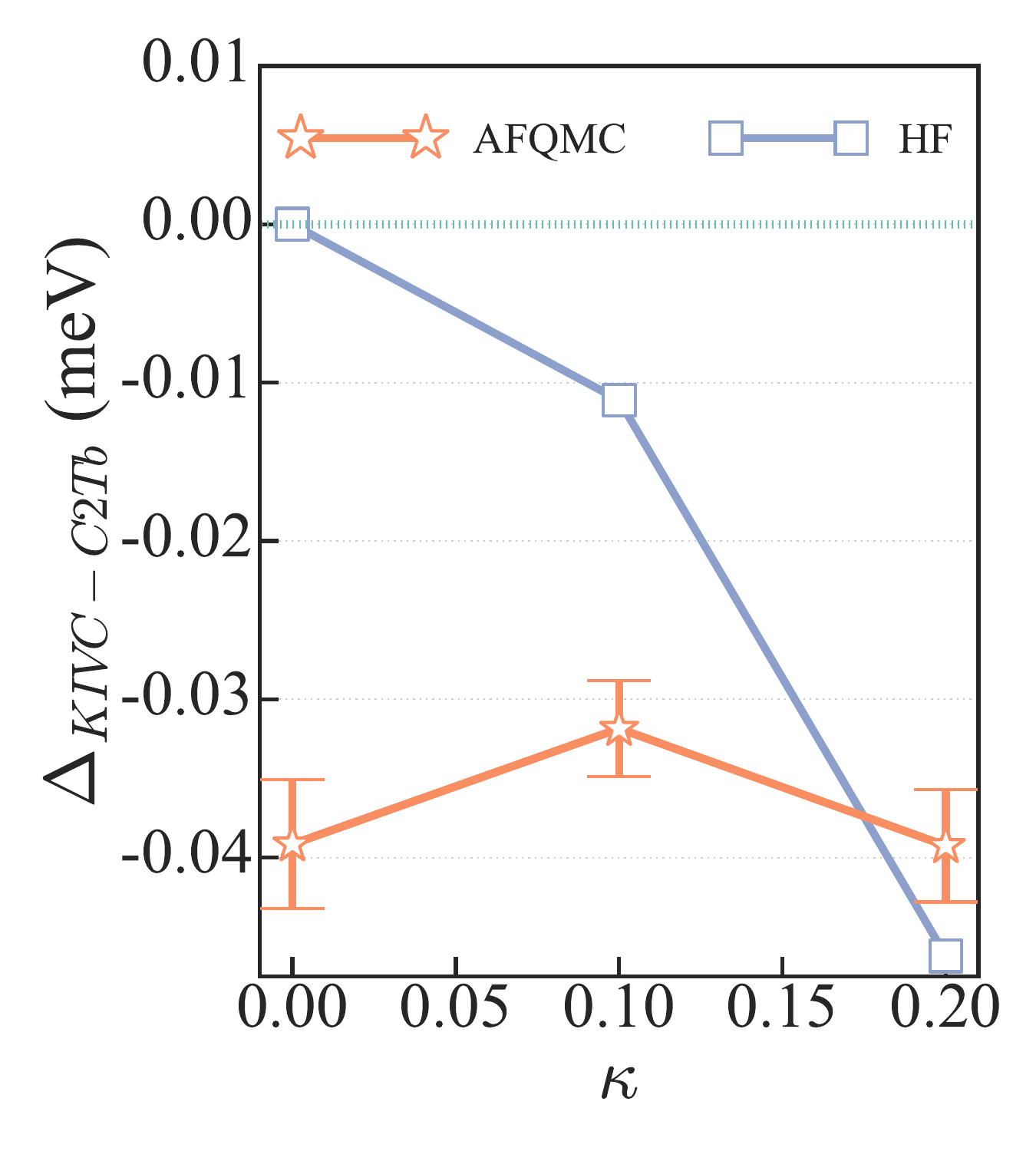}
\caption{
Energy difference per electron between the ground states in K-IVC and C$_2$T-symmetry-breaking manifolds at different $\kappa$ and $6\times 6$ lattice for $n_f=4$ fillings (left) and $n_f=2$ fillings (right). AFQMC proceeds with the corresponding HF state as trial/initial wave functions. AFQMC with Metropolis release is performed to eliminate minor constraint bias. }
\label{Fig.Energy_KIVC_C$_{2}$T_Kappa_CN+HF}
\end{figure}

\section{\label{sec:APPENDIX_order_parameter}order parameter}
In AFQMC we can use the Hellman-Feynman theorem
to calculate an order parameter $\langle \hat{O} \rangle$ by taking the derivative of energy over $\lambda$,
\cite{Absence_of_Superconductivity_MingPu}:
\begin{equation}
\langle \hat{O} \rangle  = \underset{\lambda \to 0}{\lim}\frac{\mathrm{d} \langle H_{IBM} +\lambda \hat{O} \rangle }{\mathrm{d} \lambda}.
\end{equation}
For a small range of $\lambda$ around zero, we then use the numeric derivative of the energy to estimate the average of observables in the ground state.

\begin{figure*}[htbp]
\includegraphics[scale = 0.3]{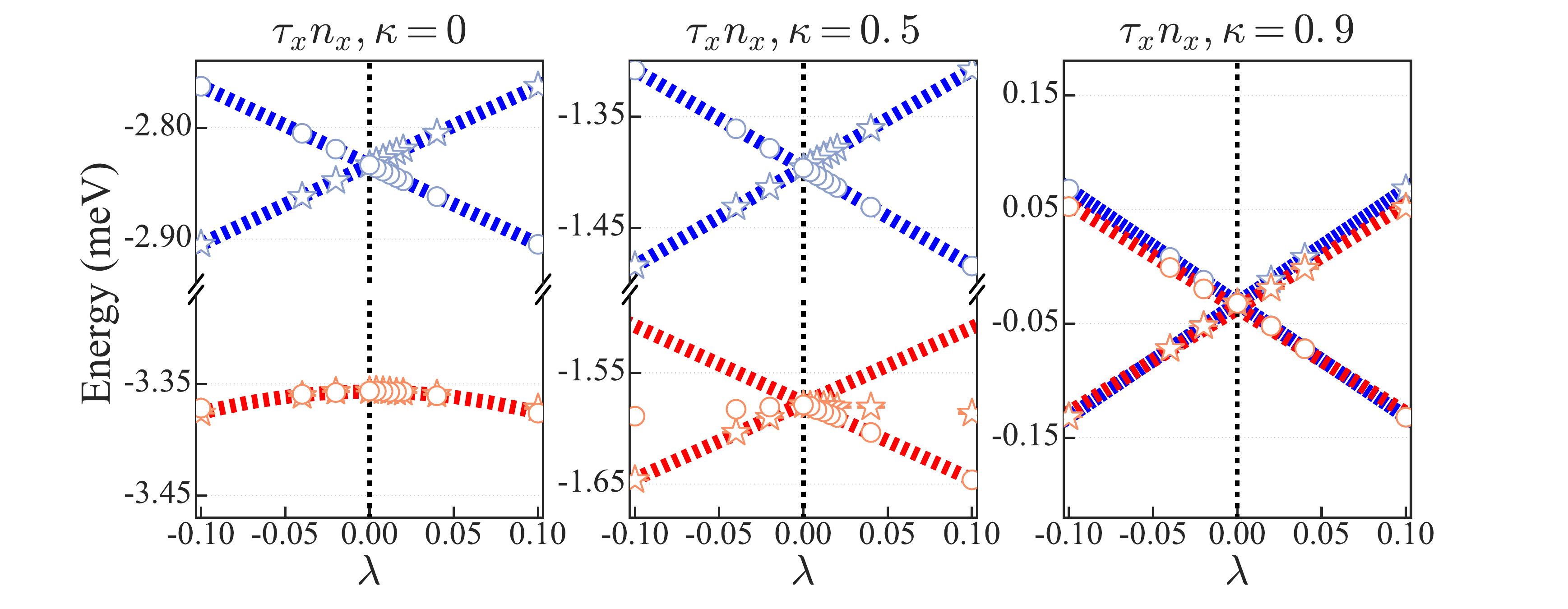} 
\label{Fig.order_parameter_details_top}
\includegraphics[scale = 0.3]{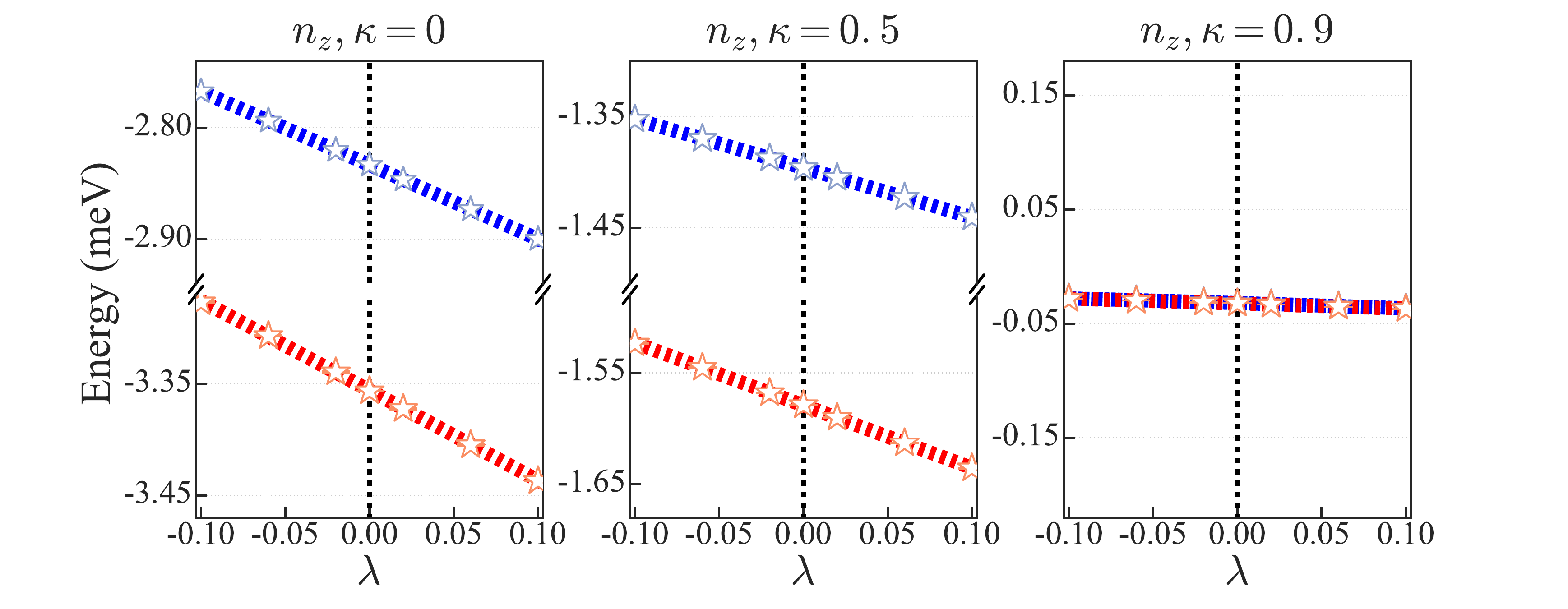}
\label{Fig.order_parameter_details_bottom}
\caption{Energy per electron versus $\lambda$ for K-IVC order parameter $\tau_x n_x$ (top)  and band occupancy $n_z$ (bottom) . HF results are presented in blue, while AFQMC results are highlighted in orange. Dashed lines/curves are linear/quadratic fitting to guide eyes. In the top figure, blue star and circle marks identify the energy of HF states that prefer $\tau_x n_x$ or $-\tau_x n_x$. The AFQMC results with corresponding HF states as trial wave functions are denoted by orange stars and circle marks. The derivative along $\lambda$ indicates the preference of the corresponding order.}
\label{Fig.order_parameter_details}
\end{figure*}

In Fig.~\ref{Fig.order_parameter_details}, we show more details about our calculations for K-IVC order parameter $\tau_x n_x$ and band occupancy $n_z$. To detect the K-IVC order, there is a two-fold degeneracy of the K-IVC pairs that favor $\tau_x n_x$ and $-\tau_x n_x$ (i.e., a direct estimation of $\tau_x n_x$ in the ground state without breaking the symmetry will always be zero). As presented in the top figures of Fig.~\ref{Fig.order_parameter_details}, adding a positive $\lambda$ will favor one of them to get lower energy and disfavor the other, which gives higher energy. We can produce a HF solution for one vs the other by imposing a corresponding initial density matrix. For $\kappa=0$ (i.e., in the metallic phase), AFQMC with different HF trial wave functions gives similar energy with zero derivatives around $\lambda=0$, supporting a loss of K-IVC order. As $\kappa$ increases to the insulating phase, the linear behavior and the increase of derivatives in AFQMC indicate the appearance of K-IVC order in the ground state. There is no such degeneracy in the calculation of band occupancy $n_z$, as the top and bottom bands have a non-zero gap. We observed a consistent linear behavior across $\lambda = 0$, and the derivatives decrease as $\kappa$ increases, which indicates a reduction of the gap between the top and bottom band.

\section{\label{sec:APPENDIX_parameter}
The choice of the permittivity $\epsilon$}

We explore the effects of the relative permittivity $\epsilon$. In our study, we chose $\epsilon \approx 10$. To briefly verify the effects of different $\epsilon$ in determining the ground state, we present a direct comparison of two $\epsilon$ in terms of charge gap at $n_f=2$ filling in Fig.~\ref{Fig.chargeGap_HFfilling_epsilon}. Our results show that by choosing a smaller $\epsilon=5$, the charge gap is almost scaled by two, but the metal-insulator transition is still maintained. 

These tests are obviously limited in scope and cannot rule out 
the possibility that some choice of $\epsilon$ 
might qualitatively change 
the ground state. 

\begin{figure}[htbp] 
\includegraphics[scale = 0.19]{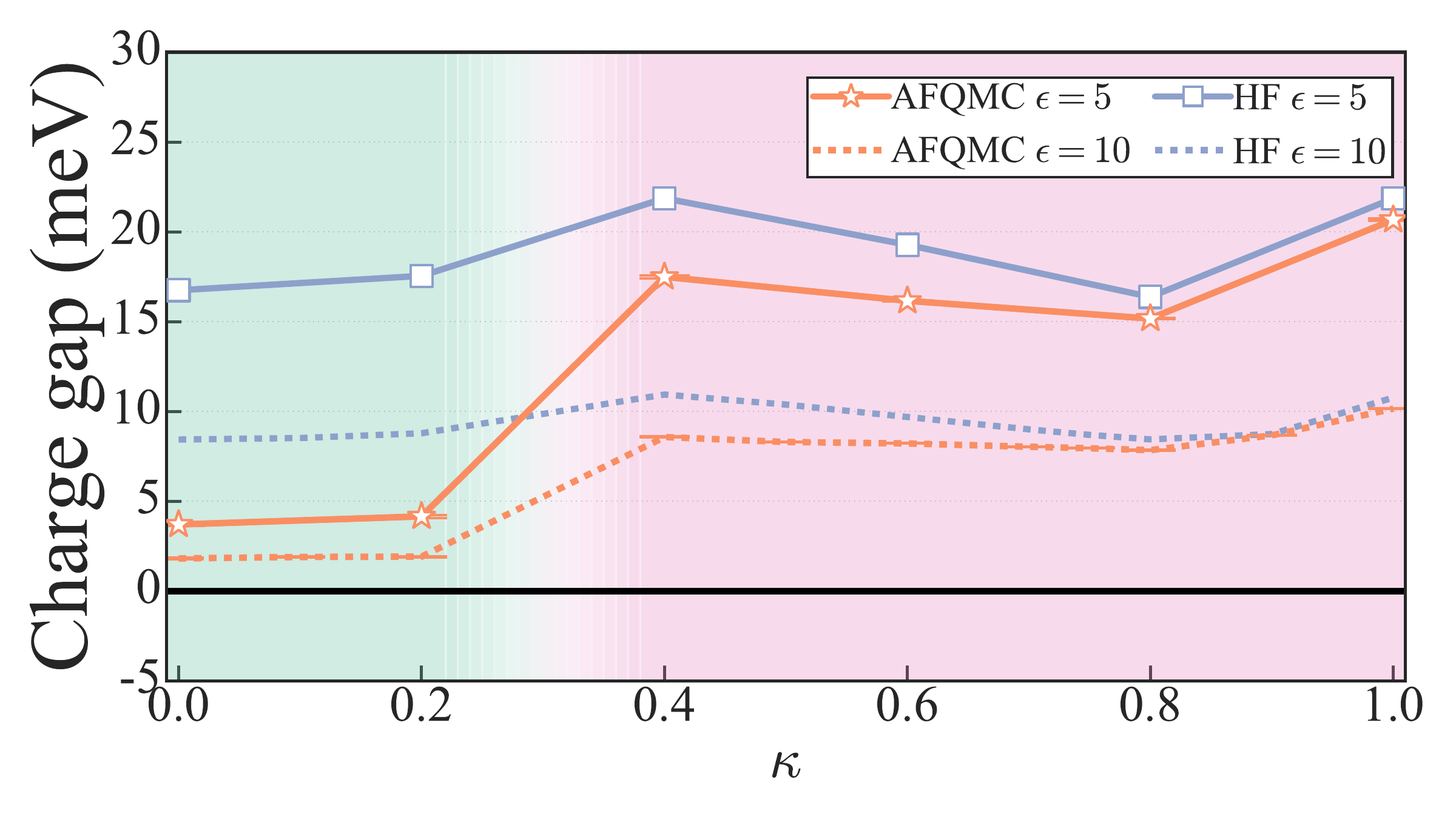} 
\caption{Charge gap for $n_f=2$ fillings versus $\kappa$ for different choice of $\epsilon$. Charge gaps are calculated by measuring the energy required to add or remove an electron from the system. AFQMC proceeds with the corresponding HF state as the trial wave function, the HF charge gap of which is illustrated in the figure. Shaded pink and green backgrounds separate the metallic/semi-metallic and insulating phases. }
\end{figure}
\label{Fig.chargeGap_HFfilling_epsilon}
\end{document}